\documentclass[10pt]{article}

\usepackage[top=20mm,bottom=25mm,left=20mm,right=20mm]{geometry}
\usepackage[T1]{fontenc}
\usepackage[utf8]{inputenc}
\usepackage{mathptmx}
\usepackage{amsmath}
\usepackage{bm}
\providecommand{\symbf}[1]{\bm{#1}}
\usepackage{graphicx}
\usepackage{booktabs}
\usepackage{multirow}
\usepackage{array}
\usepackage{xcolor}
\usepackage{colortbl}
\usepackage{caption}
\usepackage{subcaption}
\usepackage{float}
\usepackage[hidelinks]{hyperref}
\usepackage{microtype}
\usepackage{enumitem}
\usepackage{makecell}
\usepackage{placeins}
\usepackage{setspace}
\graphicspath{{./}{figures/}}
\definecolor{rowgray}{gray}{0.93}
\definecolor{rowbest}{RGB}{198,239,206}
\definecolor{rowwrst}{RGB}{255,199,206}
\definecolor{phgray}{gray}{0.96}
\makeatletter
\renewcommand\section{\@startsection{section}{1}{\z@}%
  {-10pt plus -3pt minus -2pt}{4pt plus 2pt}%
  {\normalfont\normalsize\bfseries\MakeUppercase}}
\renewcommand\subsection{\@startsection{subsection}{2}{\z@}%
  {-8pt plus -2pt minus -2pt}{3pt}%
  {\normalfont\normalsize\itshape}}
\makeatother
\setlist[itemize]{noitemsep,topsep=2pt,leftmargin=14pt}
\newcommand{\pst}{\ensuremath{P_\mathrm{st}}}
\newcommand{\pscr}{\ensuremath{P_{\mathrm{scr},20}}}

\newcolumntype{C}[1]{>{\centering\arraybackslash}p{#1}}
\newcolumntype{L}[1]{>{\raggedright\arraybackslash}p{#1}}

\newcommand{\SMnomen}{S1}    
\newcommand{\SMlcl}{S2}      
\newcommand{\SMobj}{S3}      
\newcommand{\SMrng}{S4}      
\newcommand{\SMproof}{S5}    
\newcommand{\SMtime}{S10}    

\begin{document}

\begin{center}
  {\large\bfseries STATCOM-Based Flicker Mitigation in Electric Arc Furnace
  Systems Using DQ-PI and Lyapunov-Based Nonlinear Control}\\[8pt]

  {\normalsize
   Vikram Chekuri$^{1}$,\quad
   Narapareddy Ramarao$^{1}$,\quad
   Prashanth N\,A$^{1}$}\\[4pt]

  {\small$^{1}$Department of Electrical and Electronics Engineering,\\
   BMS Institute of Technology and Management,\\
   Bengaluru, Karnataka, India\\
   Affiliated to Visvesvaraya Technological University, Belagavi, Karnataka, India\\[2pt]
   Corresponding author: Vikram Chekuri (e-mail: vikramchekuri@bmsit.in)}
\end{center}

\begin{abstract}
Electric arc furnaces inject stochastic reactive-power swings that cause voltage
flicker at the point of common coupling. This paper compares two STATCOM inner
controllers for an 11~kV, 50~Hz furnace feeder: a split-frequency DQ-PI design
and a Lyapunov-based nonlinear design whose integral state is retained in the
storage function, giving global asymptotic stability of the nominal unsaturated
current-error subsystem. Gains for both controllers were tuned by a frozen
metaheuristic campaign and then evaluated over 600~s with an IEC~61000-4-15
digital flickermeter. A post-campaign audit showed that the stochastic arc
realization used for the original comparison was periodic, so the two selected
gain sets were frozen without retuning or reselection and evaluated on ten
independent stochastic realizations that had been specified and archived before the
STO-ENS validation campaign, together with a paired uncompensated case for each
realization. Both controllers reduce the mean short-term flicker severity by about
62\% relative to the STO-ENS uncompensated baseline, and no statistically
detectable difference in mean flicker severity is found between them. They
separate instead in realization sensitivity: the Lyapunov standard deviation
across arcs is roughly one third that of DQ-PI. Controller ordering inferred
from a single stochastic trajectory is therefore not a reliable measure of
stochastic robustness, and ensemble validation is necessary.
\end{abstract}


\section*{Nomenclature}
\begin{description}[leftmargin=4.6em,style=nextline,itemsep=0pt,parsep=0pt]
\item[$\pst{}$] Short-term flicker severity over a 600~s window (IEC 61000-4-15).
\item[$\pscr{}$] Screening flicker index over a 20~s window; not an IEC-compliant $\pst{}$.
\item[$J_\mathrm{tot}$] Scalar tuning objective minimised during gain optimization.
\item[$k_{1,i},\,k_{2,i},\,k_{i,i}$] Linear, nonlinear and integral gains of the Lyapunov current law.
\item[$n_x$] Smooth-sign nonlinear current-correction term, $x\in\{d,q\}$.
\item[$\sigma_x,\,\phi_i,\,\alpha$] Boundary-layer sign function, its width, and the sublinear exponent.
\item[$\xi_{ix}$] Integral state of the current-error loop.
\item[$\tilde d_x$] Current-loop disturbance scaled by the interface inductance $L_g$.
\item[$V(\symbf e,\symbf\xi)$] Storage function used in the stability analysis.
\item[$K_\mathrm{droop}$] Voltage-droop coefficient of the Lyapunov outer loop.
\item[$\sigma_n,\,\tau_n$] Standard deviation and correlation time of the arc-length perturbation.
\end{description}

\noindent Remaining symbols are defined where they first appear; a complete list
is given in Supplementary Section~\SMnomen.

\section{Introduction}
The electric arc furnace (EAF) is among the most disruptive loads on an
industrial distribution network. It melts scrap with high-current arcs whose
length changes within milliseconds, so the arc impedance varies rapidly over a
wide range as arc length, extinction and re-ignition conditions change. The
resulting active- and reactive-power changes appear at the PCC as voltage
modulation, and because human
visual sensitivity peaks near 8.8~Hz, the EAF fluctuations in the 1--25~Hz band
are perceived as lamp flicker~\cite{ref_iec415}. Many neighbouring loads, such as
variable-speed drives and PLC-controlled production lines, are sensitive to this
disturbance.

The approximate PCC voltage change produced by a load power change is given by
equation~\eqref{eq:pcc-dv},
\begin{equation}
  \Delta V \approx \frac{R\,\Delta P + X\,\Delta Q}{V}.
  \label{eq:pcc-dv}
\end{equation}
For a predominantly inductive grid with a high reactance-to-resistance ratio ($X/R \gg 1$),
equation~\eqref{eq:pcc-dv} reduces to equation~\eqref{eq:pcc-dvx},
\begin{equation}
  \Delta V \approx \frac{X\,\Delta Q}{V}.
  \label{eq:pcc-dvx}
\end{equation}
Reactive-power fluctuation is therefore the main cause of EAF flicker, and the
natural compensation target is to drive the source-side reactive power to zero.
With the STATCOM reactive power signed according to the PCC network convention,
this is written as equation~\eqref{eq:qnull},
\begin{equation}
  Q_\mathrm{source} = Q_\mathrm{EAF} + Q_\mathrm{STATCOM} \rightarrow 0,
  \label{eq:qnull}
\end{equation}
where $Q_\mathrm{STATCOM}$ is signed according to the PCC network convention, so a
capacitive STATCOM response has a sign opposite to the furnace reactive demand and
reduces the source-side residual reactive power. In the time-domain plots of
Section~\ref{sec:results}, $Q_\mathrm{STATCOM}$ is shown as the injected
compensating component; it therefore follows the EAF reactive-power demand in
magnitude while its physical action cancels the furnace demand, leaving
$Q_\mathrm{source}$ close to zero.
Because the transformer, LCL filter, and converter interface also exchange
reactive power, the measured signals do not satisfy this two-term balance
exactly; the small interface residual is retained in all reported quantities
rather than being removed from the controller performance claim.
A static synchronous compensator (STATCOM) is well suited to EAF compensation because its current-controlled VSC gives a fast reactive-current response and retains stronger compensation capability during voltage depression than an impedance-based SVC. It also avoids the switching and resonance limitations of mechanically switched capacitor banks~\cite{ref_virtanen,ref_orcajo,ref_hussen}.

\subsection{Related Work and Research Gaps}
EAF flicker mitigation has been studied with both classical and converter-based
compensators, and converter schemes have a long track record. Larsson and
Poumar\`ede showed early that a high-bandwidth STATCOM strongly reduces EAF
flicker under sinusoidal and random arc-length modulation~\cite{ref_larsson}, and
Garc\'ia-Cerrada~\textit{et~al.} found a shunt PWM voltage-source inverter to
outperform a thyristor-controlled reactor on real furnace
data~\cite{ref_garcia}. Later work adopted nonlinear control, with Yazdani, Crow
and Guo proposing a nonlinear STATCOM controller against conventional
PI~\cite{ref_yazdani}, followed by field-scale validation of an industrial
100-MVA VSC-STATCOM~\cite{ref_morati} and selective interharmonic compensation
for light flicker~\cite{ref_feola}.
Related converter studies compared conventional and energy-storage
STATCOMs~\cite{ref_virtanen} and improved SVC control with a fast open-loop
path~\cite{ref_cernan}, and optimization-assisted tuning has recently reached EAF
duty through online SPEA-II tuning of a cascaded STATCOM~\cite{ref_liu_spea}. Outside EAF duty, optimization-based
STATCOM studies have addressed grid-code compliance and placement/parameter
setting~\cite{ref_kamel,ref_zhang}, while recent reviews of EAF simulation and numerical modelling survey how arc
behaviour is represented across process and electrical studies~\cite{ref_makki}.
Fast reactive compensation, nonlinear control, and optimization have thus each
been applied to EAF or closely related STATCOM problems, so the contribution here
is not the use of a STATCOM or an optimizer in isolation.

EAF modelling has likewise moved from deterministic resistance laws toward
stochastic and measurement-informed descriptions: three-phase time-domain models
with stochastic variation and IEC flickermeter evaluation~\cite{ref_teklic},
measurement-based time-varying models including active-power and harmonic
behaviour~\cite{ref_babaei}, and field measurements showing the practical spread
of flicker severity~\cite{ref_tlili}. These motivate using both a repeatable
benchmark and a reproducible stochastic case here.

Several nonlinear controllers provide large-signal or energy-based stability
arguments for converter systems~\cite{ref_baros,ref_jaipra,ref_jin,ref_milasi},
as do robust, disturbance-rejection, predictive and learning-based
methods~\cite{ref_hashem,ref_ahmed,ref_zhou,ref_gong,ref_xiao_mpc}, while
distribution-level DSTATCOM studies establish the importance of reference
extraction and fast compensation under changing
loads~\cite{ref_badoni,ref_notch,ref_arya_dg}.

The converter--grid interface matters as well. An LCL filter meets harmonic
requirements with smaller passive components than a single-inductor
filter~\cite{ref_chan}; active damping and gain selection for three-phase
LCL-filtered VSIs have been studied in~\cite{ref_ghoshal_ad,ref_ghoshal_ctrl};
and SRF-PLL dynamics become important when the grid interface is
weak~\cite{ref_venkat}. These results support the LCL and PLL choices used here
but do not themselves address EAF flicker-band reactive-power compensation.

That converter-based compensation reduces EAF flicker is therefore well
established, and the contribution here is not this fact but a set of narrower
open questions. First, EAF-specific nonlinear STATCOM control
exists~\cite{ref_yazdani}, but a direct long-duration comparison of a practical
split-frequency DQ-PI architecture and a Lyapunov-based nonlinear architecture on
the same converter under common plant and converter constraints is still limited. Second, stochastic and field EAF
variability are well documented~\cite{ref_teklic,ref_babaei,ref_tlili}, but the
principal methodological gap is not the absence of repeatable stochastic
training: it is the lack of a clear separation between controlled training and
independent ensemble validation. Tuning and assessment are commonly carried out
on the same stochastic realization, so optimizer repeatability and
disturbance-realization robustness are not distinguished.
Third,
optimization has now been applied directly to EAF flicker compensation
\cite{ref_liu_spea}, but multi-optimizer gain generation combined with a
separate search surrogate, 20~s post-optimization flicker screening, and 600~s
IEC flickermeter evaluation is not commonly reported. The present work is framed
around this controlled comparison; an optimizer-level statistical comparison is
outside its scope.

\subsection{Contributions}
The study builds a common STATCOM--EAF benchmark---a 500~kVA LCL-filtered
converter on an 11~kV grid (5~MVA short-circuit capacity, $X/R=7$), a repeatable nonlinear arc-voltage benchmark,
and a stochastic Cassie--Mayr arc with reproducible correlated arc-length
perturbations---and uses it to evaluate the two control architectures on a common
plant under a common evaluation protocol. A Lyapunov-based STATCOM controller with
a nonlinear inner current law,
supported by a nominal Lyapunov stability analysis of its unsaturated current-error
subsystem, is applied to EAF flicker mitigation and compared with a
split-frequency DQ-PI controller on the same converter. Six gains of each
controller are tuned by five metaheuristic optimizers against one common
power-quality objective with mode-aware search bounds, and every optimized gain
set then enters the three-stage evaluation pipeline: 200 optimized gain sets are screened with
a 20~s flicker index, 20 task-wise sets are selected, and each selected set is
evaluated over 600~s with an IEC~61000-4-15 digital flickermeter. The resulting
long-duration flicker behaviour separates mean performance from realization
sensitivity: the STO-SR single-trajectory evaluation appeared to show a
regime-dependent controller ranking, whereas STO-ENS validation finds
a Lyapunov advantage under the deterministic arc but no statistically detectable
mean difference under independent stochastic arc realizations, the controllers differing
instead in arc-to-arc dispersion. Time-domain
metrics---source reactive-power fluctuation, dc-link voltage deviation, PCC
voltage THD, $d$- and $q$-axis current-tracking error, and converter
saturation---explain the practical trade-offs of the two controllers on a common
basis.

\section{System Configuration}
\begin{figure}[!htbp]
  \centering
  \includegraphics[width=0.82\linewidth]{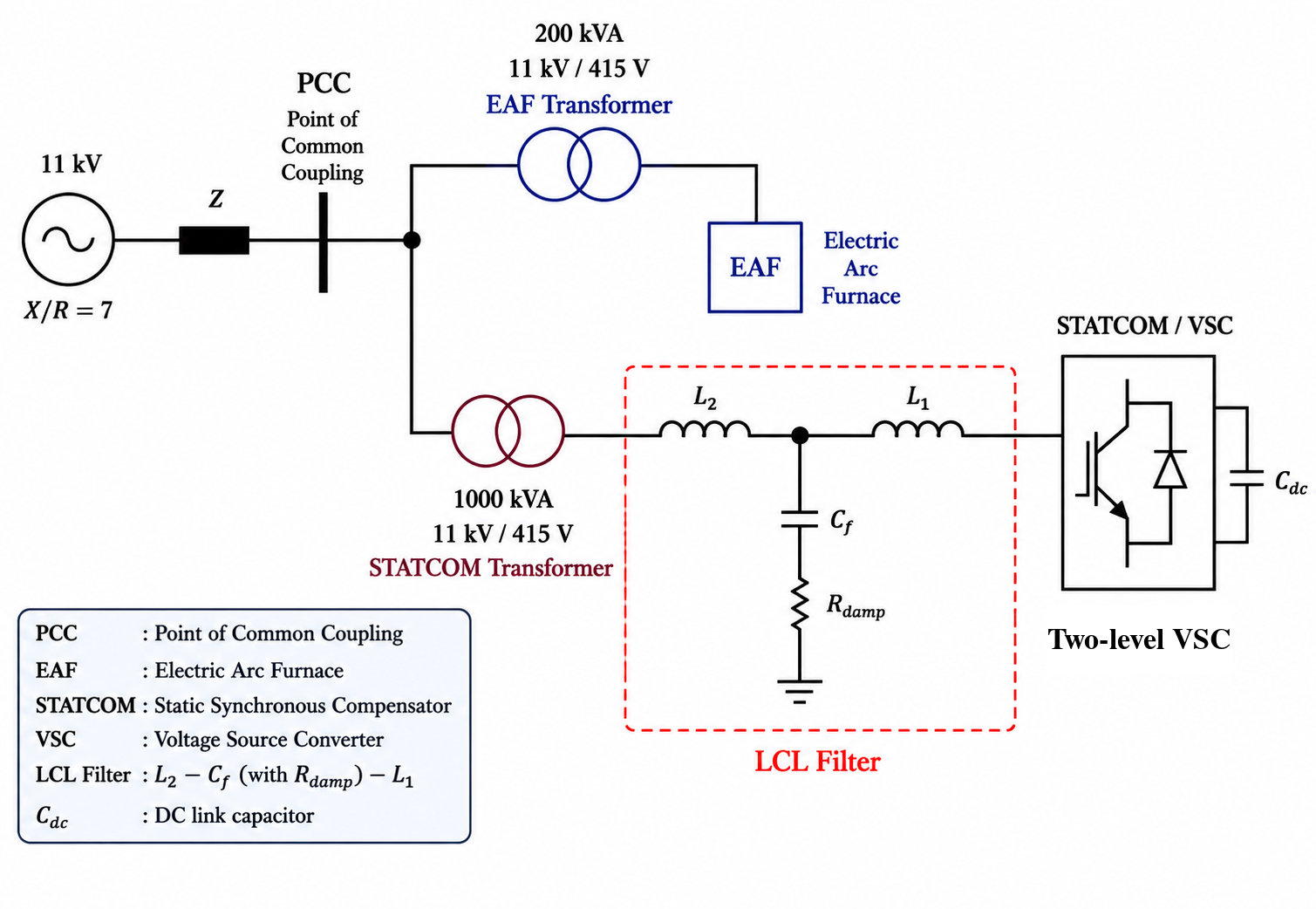}
  \caption{Single-line configuration of 11~kV grid, electric arc furnace
           load and LCL-filtered two-level VSC-STATCOM.}
  \label{fig:sysconf}
\end{figure}

An 11~kV, 50~Hz grid supplying an EAF through a furnace transformer is
shown in Fig.~\ref{fig:sysconf}. A 500~kVA two-level VSC-STATCOM is connected
in shunt at the PCC through a 1000~kVA, 11/0.415~kV Dyn1 coupling transformer
and an LCL filter, while the 200~kVA furnace is supplied through a separate
11/0.415~kV Dyn1 transformer. Both transformers are non-ideal, modelled with
$0.005$~pu winding resistance and $0.03$~pu leakage reactance per winding and a
$500$~pu magnetizing branch, with saturation disabled. The grid short-circuit capacity is 5~MVA with
$X/R=7$, representing a finite-strength, predominantly inductive supply
in which furnace reactive-power variations produce observable PCC voltage
modulation. The converter is supplied by a 1200~V, 70~mF dc link and operates at 
a switching frequency of 5~kHz with a simulation time step of 10~$\mu$s. With
$m_\mathrm{max}=0.95$, the available converter phase-peak voltage is 570~V.
The converter ratings, physical current and voltage ceilings, plant model, and
operating conditions are identical for both controllers; controller-specific
reference limiting, anti-windup, and protection logic are treated as part of each
implemented control architecture, so the comparison is between two complete
architectures on a common plant.

The rated converter current follows from the three-phase apparent-power relation
of equation~\eqref{eq:irated},
\begin{equation}
  I_\mathrm{rated}=\frac{S_\mathrm{rated}}{\sqrt{3}\,V_{LL}}
  =\frac{500\times10^{3}}{\sqrt{3}\times 415}\approx 695.6~\mathrm{A},
  \label{eq:irated}
\end{equation}
and is the normalization base for every current limit and for the
current-tracking penalty term. The available pulse-width-modulation (PWM) phase-peak voltage is
$V_\mathrm{pk,max}=m_\mathrm{max}V_\mathrm{dc}/2=0.95\times1200/2=570$~V, which
sets the voltage-saturation boundary used in equation~\eqref{eq:vsat}.

\subsection{LCL Filter Design}
The converter connects to the grid through an LCL filter designed by the
standard procedure of~\cite{ref_chan}, with $L_1=0.22$~mH, $L_2=0.20$~mH,
$C_f=280~\mu$F ($3.0\%$ of the base capacitance) and a passive damping resistor
$R_d=0.2~\Omega$. The resulting resonance is approximately $930$~Hz, satisfying
$10f_\mathrm{line}<f_\mathrm{res}<f_\mathrm{sw}/2$, and the filter attenuates the
5~kHz switching component by $34.8$~dB, above the adopted $30$~dB criterion. The
full sizing sequence, the third-order filter transfer function, the attenuation
calculation and the damping choice are given in Supplementary Section~\SMlcl.

\section{Deterministic and Stochastic EAF Modeling}
The arc voltage of an EAF is a nonlinear function of the arc current and shows a
hysteretic I--V loop: a high re-ignition voltage at each current zero crossing, a
falling-then-flat conduction region, and an extinction segment. As the electrodes
move and the scrap melts, the arc length---and therefore the arc-voltage
magnitude---changes continuously, with two effects. First, the
fundamental-frequency reactive demand fluctuates and modulates the PCC voltage
envelope, as described by equation~\eqref{eq:pcc-dv}. Second, the nonlinear I--V
relationship injects harmonic and inter-harmonic currents that distort the
waveform. The perceptual severity of the envelope modulation is measured by the
short-term flicker index \pst{}, computed by the IEC~61000-4-15 flickermeter,
which weights the fluctuation by the lamp--eye--brain response and aggregates it
over a 10-minute window~\cite{ref_iec415}; $\pst{}=1.0$ is used here as a
commonly adopted perceptibility benchmark, following the guidance framework of
IEC~TR~61000-3-7~\cite{ref_iec37}. In the present electromagnetic-transient
(EMT) study, controller gains are optimized against the common power-quality
objective of Eq.~\eqref{eq:J} over a $0.50$--$1.00$~s window; all optimized gain
sets are then screened with a
20~s post-optimization flicker index, and the task-wise selected gain sets are
evaluated through full 600~s EMT simulations with the IEC flickermeter
(Section~\ref{sec:val}). During the optimization campaign the stochastic disturbance was held to the
single reproducible realization STO-SR, identical in every run and for both
controllers, so that optimizer variability was not confounded by arc-to-arc
variation. The final controller comparison is different: the selected gains are
frozen, without retuning or reselection, and evaluated across the ten
pre-specified realizations of STO-ENS, each with its paired uncompensated case
(Section~\ref{sec:corrected}). No certified network-compliance measurement is claimed. Because the
high-sensitivity flicker region emphasized here ($7$--$12$~Hz) lies well below the
line frequency, the compensator must reproduce and oppose the reactive-power
fluctuation in near real time, which is where a VSC-STATCOM outperforms slower
compensators.

The EAF model has two operating modes, which the three experimental conditions of
Table~\ref{tab:eafcases} draw on. The deterministic mode is a repeatable
nonlinear arc-voltage benchmark (DET). The stochastic mode retains dynamic
Cassie--Mayr conductance states and adds a seeded, reproducible arc-length
perturbation; it is used in two distinct roles, as the single fixed-seed
realization STO-SR and as the independent ensemble STO-ENS. The I--V
characteristics and waveforms are compared in Fig.~\ref{fig:eafmodes}: the
deterministic model gives a smooth, repeatable I--V locus, while the stochastic
model shows dispersed loops and irregular waveforms from seeded arc-length
fluctuation, conductance dynamics, phase imbalance, and arc
extinction/re-ignition.

\subsection{Deterministic Nonlinear Arc-Voltage Benchmark}
The deterministic model is intentionally simple and repeatable. It is not a
dynamic Cassie--Mayr conductance model; instead, it uses a smooth nonlinear
arc-voltage law so that the two controllers see exactly the same periodic EAF
excitation. In the deterministic model, the external flicker signal is a sinusoid of
$50.26$~rad/s, i.e., approximately 8.0~Hz, close to the frequency range of high
flicker sensitivity. The input $f[k]$ is first low-pass filtered to represent a
finite electrode/arc response, as in equation~\eqref{eq:flp},
\begin{equation}
  f_\mathrm{lp}[k]=(1-a)\,f_\mathrm{lp}[k-1]+a\,\mathrm{clip}(f[k],-1,1),
  \quad a=\frac{T_s}{\tau+T_s},
  \label{eq:flp}
\end{equation}
with $T_s=10~\mu$s and $\tau=15$~ms. The modulation depth weakens as the arc
current increases, as written in equation~\eqref{eq:meff},
\begin{equation}
  m_\mathrm{eff}=\frac{m_0}{1+0.4\,|i|/300},\qquad m_0=0.1,
  \label{eq:meff}
\end{equation}
and the smooth sign function $s(i)=\tanh(i/I_s)$ with $I_s=40$~A avoids a hard
switch at the current zero. With
$v_t=V_\mathrm{at0}(1+m_\mathrm{eff}f_\mathrm{lp})$ and
$V_\mathrm{at0}=220$~V, the intrinsic arc voltage is given by equation~\eqref{eq:detarc},
\begin{equation}
  v=s(i)\!\left(v_t+\frac{C}{D+|i|+\varepsilon}\right)+r_a\,i,
  \label{eq:detarc}
\end{equation}
where $C=1.5\times10^4$~V\,A, $D=7000$~A,
$\varepsilon=10^{-3}$~A, and $r_a=0.03~\Omega$. This model provides the
repeatable nonlinear I--V characteristic shown in Fig.~\ref{fig:eafmodes}(a); it is
used as a benchmark rather than as a claim of complete furnace physics.

\subsection{Stochastic Cassie--Mayr Arc with Reproducible Correlated Noise}
\label{sec:eaf}
The stochastic mode adds dynamic Cassie and Mayr conductance states together
with a slowly varying arc-length perturbation. The perturbation follows the
mean-reverting discrete update in equation~\eqref{eq:ou}, which is inspired by
the Ornstein--Uhlenbeck process~\cite{ref_ou},
\begin{equation}
  \eta[k]=(1-\alpha_n)\eta[k-1]+\sigma_n\sqrt{2\alpha_n}\,w[k],
  \quad \alpha_n=\frac{T_s}{\tau_n}.
  \label{eq:ou}
\end{equation}
Here $\tau_n=0.25$~s and $\sigma_n=0.015$. The implementation does
not assume a Gaussian driver. Instead, $w[k]$ is produced by a seeded
linear-congruential pseudo-random sequence and scaled by 3.46, giving an
approximately zero-mean, unit-variance bounded driver. The effective phase seeds
are $42+17p$, where $p=1,2,3$, giving 59, 76, and 93. The fixed seeds make the
stochastic trajectory exactly reproducible for all controller and optimizer
comparisons.

\begin{figure*}[!t]
\centering
\includegraphics[width=0.98\textwidth]{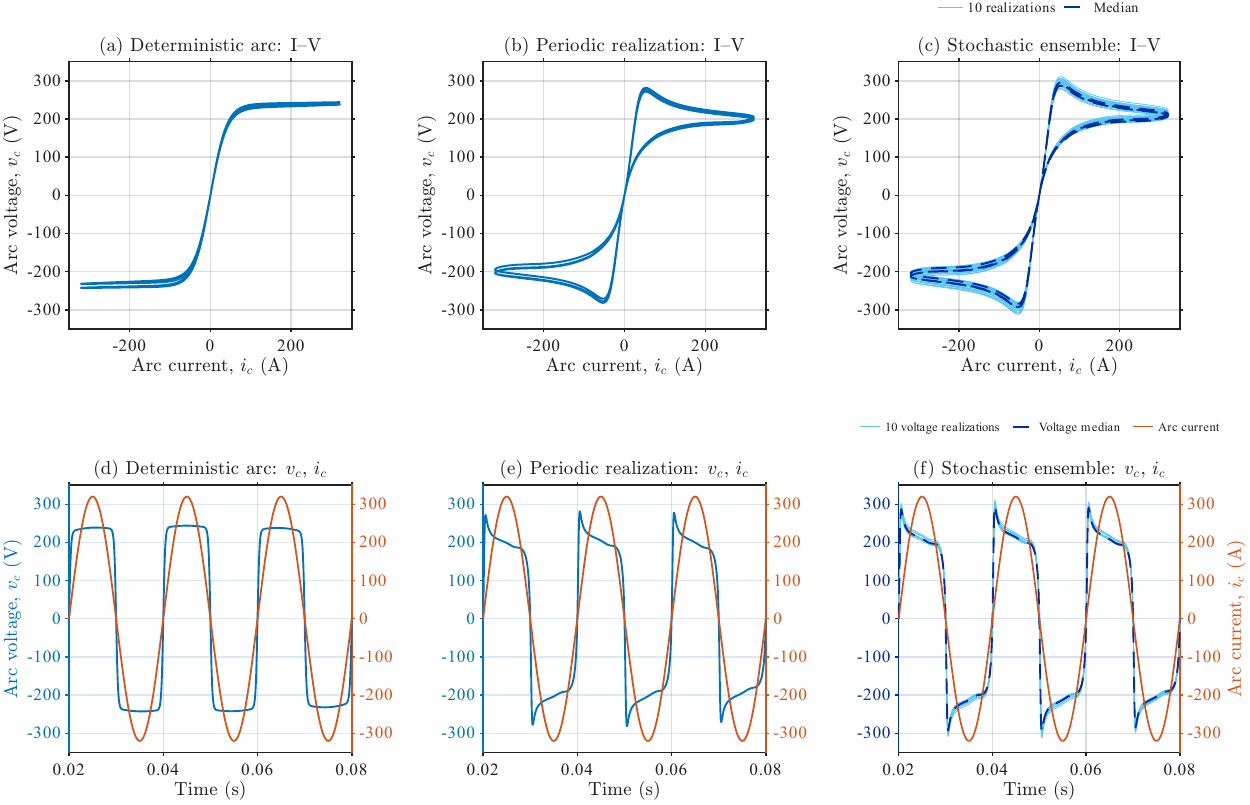}
\caption{Intrinsic EAF characteristics over three representative line cycles
under the three conditions of Table~\ref{tab:eafcases}. Panels (a)--(c) show arc
I--V trajectories and panels (d)--(f) show arc voltage and current against time,
for DET, STO-SR and STO-ENS respectively. In (c) and (f) the light-cyan traces
denote the ten STO-ENS realizations and the dashed dark-blue trace denotes their
pointwise ensemble median, a visualization statistic rather than an additional physical
realization. All six panels share the same display window $0.02$--$0.08$~s,
selected from the frozen five-cycle record; the simulation was neither shortened
nor repeated. The embedded panel titles read ``Deterministic arc'', ``Periodic
realization'' and ``Stochastic ensemble''; the middle column is the fixed-seed
stochastic-mode realization STO-SR, whose archived trajectory was found to be
periodic, and the right column is the stochastic ensemble STO-ENS generated with
the exact-integer sampler from the pre-specified bases $5001$, $5054$, $5107$,
$5160$, $5213$, $5266$, $5319$, $5372$, $5425$ and $5478$. The deterministic
nonlinear arc physics is identical in all three columns, so the central I--V
characteristic is expected to agree; stochasticity appears as
realization-to-realization dispersion rather than as a change of characteristic.
In (d)--(f) the arc voltage $v_c$ is read against the left axis and the arc
current $i_c$ against the right axis. Per-realization dispersion is resolved in
Supplementary Fig.~S4.}
\label{fig:eafmodes}
\end{figure*}

The correlated perturbation enters the normalized arc length together with the
filtered 8~Hz modulation and a small phase offset, as in equation~\eqref{eq:larc},
\begin{equation}
  L_\mathrm{arc}[k]=\mathrm{clip}\!\left(
  L_0+\Delta L_p+m_\mathrm{flicker}f_\mathrm{lp}[k]+\eta[k],\;0.6,\;1.4\right),
  \label{eq:larc}
\end{equation}
where $L_0=1$, $m_\mathrm{flicker}=0.065$, and
$\Delta L_p\in\{0,0.01,0.02\}$. The Cassie voltage and Mayr power levels then
follow directly from the arc length through equation~\eqref{eq:cmlevels},
\begin{equation}
  V_c=V_{c0}L_\mathrm{arc},\qquad P_m=P_{m0}L_\mathrm{arc},
  \quad V_{c0}=200~\mathrm{V},\;P_{m0}=5\times10^4~\mathrm{W}.
  \label{eq:cmlevels}
\end{equation}
The two conductance states follow the classical Cassie and Mayr
formulations~\cite{ref_cassie,ref_mayr}, in the combined black-box form used for
arc-furnace flicker studies~\cite{ref_teklic}, and evolve according to
equation~\eqref{eq:cm},
\begin{equation}
  \frac{dg_c}{dt}=\frac{g_c}{\tau_c}\!\left[\!\left(\frac{v_\mathrm{arc}}{V_c}\right)^2-1\right],
  \qquad
  \frac{dg_m}{dt}=\frac{g_m}{\tau_m}\!\left(\frac{P_\mathrm{arc}}{P_m}-1\right),
  \label{eq:cm}
\end{equation}
with $\tau_c=0.5$~ms and $\tau_m=0.3$~ms. During active conduction, the
intrinsic arc voltage follows equation~\eqref{eq:stoarcv},
\begin{equation}
  v_\mathrm{arc}=\frac{i}{g_c+g_m}+r_a i,
  \label{eq:stoarcv}
\end{equation}
with $g_c$ and $g_m$ limited to $[g_\mathrm{min},g_\mathrm{max}]=[0.02,50]$~S and
$r_a=0.02~\Omega$; the final model voltage is limited to $\pm500$~V. Low-current
extinction and re-ignition are
included using $i_\mathrm{ext}=2$~A, an extinction rate coefficient
$p_\mathrm{ext}=0.003$~ms$^{-1}$, $V_\mathrm{re}=280$~V, and
$g_\mathrm{re}=0.25$~S. While the
arc conducts, if $|i|<i_\mathrm{ext}$ it extinguishes stochastically with per-step
probability $p_\mathrm{ext}\,(1-|i|/i_\mathrm{ext})\,T_s\times10^{3}$; once
extinguished, it re-ignites when $|i|/g_\mathrm{min}>V_\mathrm{re}L_\mathrm{arc}$
or $|i|>3i_\mathrm{ext}$, at which point both conductances are reset to
$g_\mathrm{re}$. Three EAF conditions are distinguished throughout this paper. The
\emph{deterministic EAF} (DET) is the
repeatable deterministic arc formulation. The \emph{fixed-seed stochastic-mode
realization} (STO-SR) is the single reproducible stochastic-mode trajectory used
during the original optimization campaign; post-campaign generator auditing
showed this archived realization to be periodic, so it is not interpreted as an
independent stochastic sample. The \emph{stochastic ensemble} (STO-ENS) denotes
independent multi-realization stochastic validation carried out after the
relevant gain selection, with the gains frozen and without retuning or
reselection. STO-ENS is therefore a methodological role rather than one fixed
dataset: the number, duration and identity of its realizations are
study-specific. In this paper STO-ENS is the ten independent long-duration
realizations of Section~\ref{sec:corrected}. STO-SR and STO-ENS use the same stochastic arc model and
differ in experimental role, not in physics. Archived source files, workbooks and
dataset fields retain the historical \texttt{STO} identifier for traceability;
the manuscript distinguishes the two roles explicitly. Thus the stochastic case
is not simply the deterministic curve with added noise: its arc length changes the
Cassie and Mayr operating levels, while conductance memory and
extinction/re-ignition create the dispersed I--V trajectories seen in
Fig.~\ref{fig:eafmodes}(c). Similar stochastic and measurement-informed EAF modelling
has been used to study flicker and power-quality behaviour in
\cite{ref_teklic,ref_babaei}. The stochastic-model parameters are listed
in Supplementary Table~S1. Under the present grid condition, this model
raises the uncompensated 600~s \pst{} from 1.812 to 2.262.

A post-campaign audit established that the sampler used during the campaign
produced a periodic rather than broadband arc-length perturbation, so that
condition is the stochastic-mode realization (STO-SR) defined above; the deterministic arc mode
is unaffected and bit-identical between implementations. The controller gains
were therefore frozen without retuning and independently re-evaluated over the
ten pre-specified realizations of the stochastic ensemble
(Section~\ref{sec:corrected}). Exact cycle length, the low-order-bit mechanism,
executed-versus-intended perturbation statistics and the seed diagnostics are
reported in Supplementary Section~\SMrng.

\begin{table}[!t]
\centering
\caption{The three EAF conditions used in this work. STO-ENS was used only for
frozen-gain validation: no gain was retuned and no champion was reselected
against it. STO-ENS denotes a methodological role rather than one fixed dataset;
in this paper it comprises ten independent long-duration realizations. Archived
identifiers containing \texttt{STO} are retained in the
source files and datasets for traceability.}
\label{tab:eafcases}
\setlength\tabcolsep{4pt}\renewcommand{\arraystretch}{1.12}\small
\begin{tabular}{llccc}
\toprule
\textbf{Label} & \textbf{Manuscript term} & \textbf{Arc realization} &
\textbf{Used for} & \textbf{Used for 600~s}\\
 & & & \textbf{tuning?} & \textbf{validation?}\\
\midrule
DET     & Deterministic EAF            & deterministic              & Yes & Yes\\
STO-SR  & Fixed-seed stochastic-mode   & one reproducible           & Yes & Reference\\
        & realization                  & fixed-seed trajectory      &     & \\
STO-ENS & Stochastic ensemble          & independent multi-         & No  & Yes, primary\\
        & (role; ten in this paper)    & realization set            &     & \\
\bottomrule
\end{tabular}
\end{table}

\section{STATCOM Control Architecture}
\label{sec:control}
\begin{figure*}[!t]
\centering
\includegraphics[width=\textwidth]{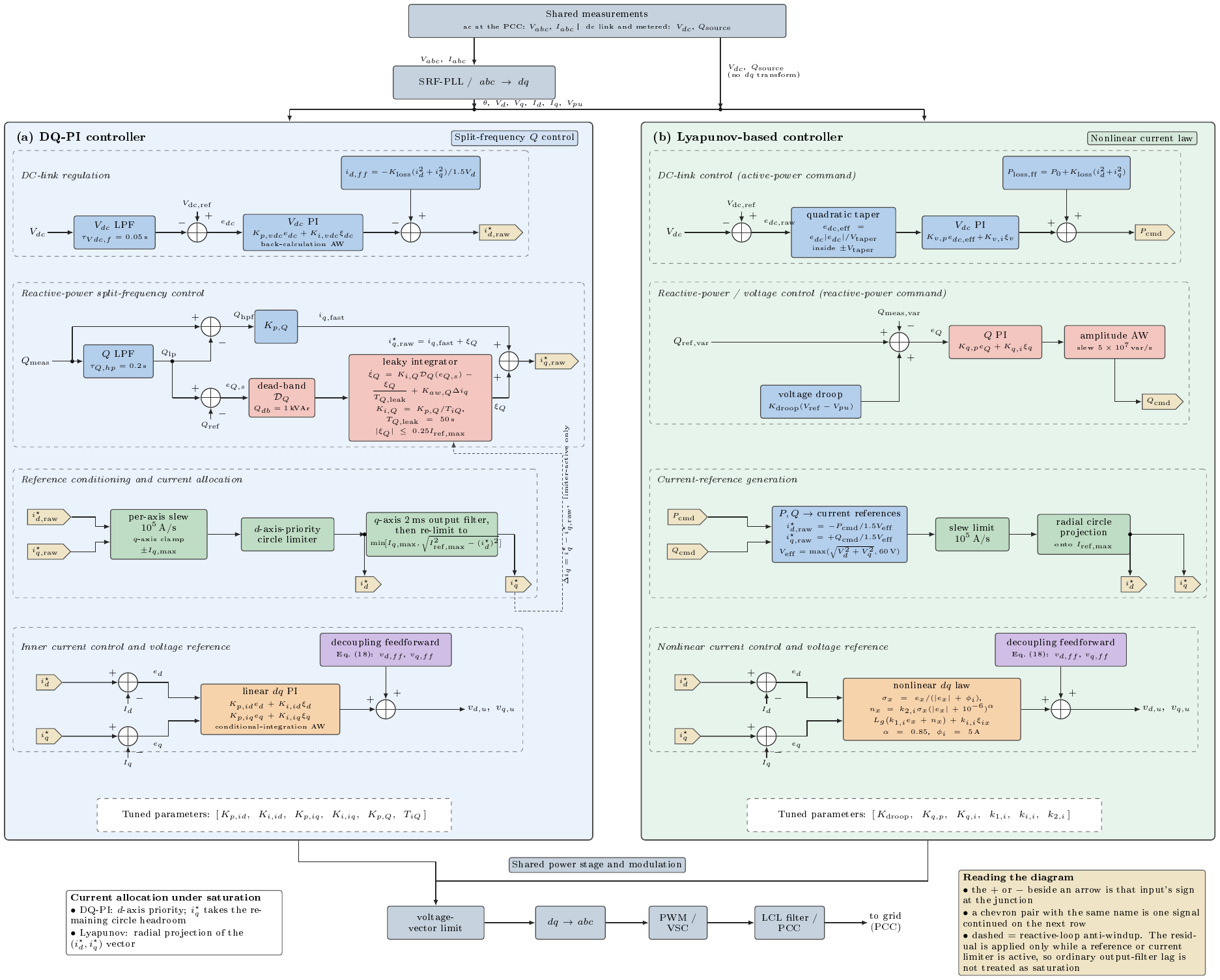}
\caption{Signal-flow architecture of the two compared STATCOM controllers. Both
branches share the measured quantities shown at the top. The PCC ac measurements
$V_{abc}$ and $I_{abc}$ pass through the SRF-PLL and $abc\rightarrow dq$
transformation, whereas $V_\mathrm{dc}$ and $Q_\mathrm{source}$ bypass that
transformation and feed the relevant outer loops directly. Both branches share
the voltage-vector limiter, $dq\rightarrow abc$ transformation, PWM/VSC stage and
LCL filter. (a) The DQ-PI controller regulates the dc link
through a filtered PI loop with $I^{2}$ loss feedforward, and splits reactive
regulation into a fast proportional path on $Q_\mathrm{hpf}$ and a slow
dead-band, leaky-integral path with conditional anti-windup; the combined
reference passes a slew limit, the $d$-axis-priority circle limiter, the 2~ms
$q$-axis output filter and the final re-limit of
equation~\eqref{eq:circle}; the anti-windup residual $\Delta i_q$ returns to the
leaky integrator (dashed), and is applied only while a reference or current
limiter is active, so ordinary output-filter lag is not treated as saturation.
(b) The Lyapunov controller forms an active-power command through a quadratic
taper, and a reactive-power command through a droop-augmented PI with limiter and
slew, converts $(P,Q)$ to current references with radial circle projection, and
closes the inner loop with the nonlinear law of
equation~\eqref{eq:lyapv}. The two designs therefore differ in current
allocation under saturation: DQ-PI clamps the $d$ axis first and gives $I_q$ the
remaining headroom, whereas the Lyapunov design projects the $(I_d^{\star},
I_q^{\star})$ vector radially. The optimizer-tuned parameter vectors are listed
explicitly in the bottom box of each panel; all other numerical parameter
values shown are fixed design values:
filter time constants, anti-windup constants, $\alpha$,
$\phi_i$, current ceilings and protection thresholds. A chevron pair carrying the
same signal name denotes one signal continued on the next row.}
\label{fig:arch}
\end{figure*}

The STATCOM is a two-level VSC operated in the synchronous $dq$ frame aligned to
the PCC voltage by a synchronous-reference-frame phase-locked loop (PLL). The PLL
regulates the normalized $q$-axis voltage error through a PI law (proportional
gain $280$, integral gain $2.5\times10^{4}$) about a $2\pi\times50$~rad/s
feedforward, with its estimated frequency clamped to $\pm20\%$; this moderate
bandwidth keeps its dynamics from interacting with the current loops at this grid
strength~\cite{ref_venkat}. Under grid-voltage orientation ($v_q\approx0$) the
instantaneous powers are $P=\tfrac{3}{2}v_d i_d$ and $Q=-\tfrac{3}{2}v_d i_q$, so
the $d$-axis current regulates dc-link energy and the $q$-axis current performs
reactive (flicker) compensation. Both controllers share an outer dc-link loop, a reactive-power loop, current
slew/circle limiters and a final voltage-vector saturation block, and differ in
how each loop is built; Fig.~\ref{fig:arch} shows the two signal paths side by
side and marks the six tuned gains of each design. The shared
voltage-vector saturation is given by equation~\eqref{eq:vsat},
\begin{equation}
  (v_d^{*},v_q^{*})=
  \begin{cases}
  \dfrac{V_\mathrm{pk,max}}{V_\mathrm{mag}}(v_{d,u},v_{q,u}), & V_\mathrm{mag}>V_\mathrm{pk,max},\\[4pt]
  (v_{d,u},v_{q,u}), & \text{otherwise},
  \end{cases}
  \label{eq:vsat}
\end{equation}
with $V_\mathrm{mag}=\sqrt{v_{d,u}^{2}+v_{q,u}^{2}}$. When the upper branch is
active, the converter-saturation penalty rises and the inner
integrators are gated. Both control laws generate limited $dq$ voltage commands, which are transformed to three-phase references and applied to the common PWM stage.

\subsection{DQ-PI Controller}
The six optimizer-tuned DQ-PI gains are
$\symbf{x}_\mathrm{DQPI}=[K_{p,id},K_{i,id},K_{p,iq},K_{i,iq},K_{p,Q},T_{iQ}]$. The dc-link
loop filters the measured dc voltage ($\tau_{Vdc,f}=0.05$~s) and combines a PI
law with a converter-loss feedforward, as in equation~\eqref{eq:idraw},
\begin{equation}
  i_{d,raw}^{*}=i_{d,ff}-\bigl(K_{p,vdc}\,e_{dc}+K_{i,vdc}\,\xi_{dc}\bigr),\qquad
  i_{d,ff}=-\frac{K_\mathrm{loss}\,(i_d^{2}+i_q^{2})}{1.5\,V_d}.
  \label{eq:idraw}
\end{equation}
Here $e_{dc}=V_\mathrm{dc,ref}-V_\mathrm{dc,f}$ is the filtered dc-voltage error,
with $V_\mathrm{dc,f}$ the low-pass-filtered measured dc voltage, and $\xi_{dc}$
is its integral. The
fixed (non-tuned) gains $K_{p,vdc}=0.992$ and $K_{i,vdc}=1.190$ follow a
margin-based design on the integrator plant $1.5V_d/(C_\mathrm{dc}V_\mathrm{dc})$:
the loop crosses over near $6$~rad/s with a $62^{\circ}$ phase margin, the PI
zero sits a decade below crossover, and the loop gain stays well below unity
across the flicker band, so the dc loop does not re-inject flicker through the
$d$-axis. The quadratic loss coefficient is $K_\mathrm{loss}=0.002$~W/A$^{2}$, the same coefficient used by the Lyapunov controller. The Lyapunov power command additionally contains a fixed 1.5~kW loss offset. In DQ-PI, the quadratic term supplies the estimated conduction-loss current directly and reduces the burden on the dc-link integrator; the dc-link integrator uses back-calculation anti-windup with $K_{aw,Vdc}=5$.
The reactive loop splits the source reactive power into a fast fluctuating
component and a slow mean component. A first-order high-pass filter gives
$Q_\mathrm{hpf}=Q_\mathrm{meas}-Q_\mathrm{lp}$ ($\tau_{Q,hp}=0.2$~s), and the
reactive current reference combines a proportional action on $Q_\mathrm{hpf}$
with an integral action on the slow reactive error, as in
equations~\eqref{eq:qdb} and~\eqref{eq:iqraw}. With
$e_{Q,s}=Q_\mathrm{lp}-Q_\mathrm{ref}$, the slow error is first passed through a
1~kVAr dead-band,
\begin{equation}
  \mathcal{D}_Q(e)=
  \begin{cases}
    0, & |e|\le Q_{db},\\
    \mathrm{sgn}(e)(|e|-Q_{db}), & |e|>Q_{db},
  \end{cases}
  \qquad Q_{db}=1~\mathrm{kVAr},
  \label{eq:qdb}
\end{equation}
and the fast and slow paths are combined as
\begin{align}
  i_{q,raw}^{*}&=K_{p,Q}\,Q_\mathrm{hpf}+\xi_Q,\qquad
  K_{i,Q}=\frac{K_{p,Q}}{T_{iQ}}, \nonumber\\
  \dot{\xi}_Q&=K_{i,Q}\,\mathcal{D}_Q(e_{Q,s})
  -\frac{\xi_Q}{T_{Q,\mathrm{leak}}}+K_{aw,Q}\,\Delta i_q,
  \label{eq:iqraw}
\end{align}
where $K_{p,Q}<0$, $T_{iQ}>0$, $T_{Q,\mathrm{leak}}=50$~s, and
$K_{aw,Q}=20$. The anti-windup residual $\Delta i_q=i_q^{*}-i_{q,raw}^{*}$ is
used only when a real reference/current limiter is active; ordinary lag of the
output filter is not treated as saturation. The slow integral is limited to
$0.25I_\mathrm{ref,max}$ (about 121.7~A). Both raw references are rate-limited (slew limit $10^5$~A/s) and the $q$-axis
reference is clamped to $\pm I_{q,\max}$ before the $d$-axis-priority circle
limiter of equation~\eqref{eq:circle},
\begin{equation}
  \text{if }\sqrt{(i_d^{*})^{2}+(i_q^{*})^{2}}>I_\mathrm{ref,max}:\;\;
  i_d^{*}\leftarrow\mathrm{sat}\bigl(i_d^{*},I_{d,\max}\bigr),\;\;
  i_q^{*}\leftarrow\mathrm{sat}\Bigl(i_q^{*},\sqrt{I_\mathrm{ref,max}^{2}-(i_d^{*})^{2}}\Bigr),
  \label{eq:circle}
\end{equation}
where $\mathrm{sat}(x,L)$ clamps $x$ to $[-L,L]$. The ceilings
$I_\mathrm{ref,max}=486.9$~A, $I_{d,\max}=173.9$~A and $I_{q,\max}=452.1$~A
($0.70$, $0.25$ and $0.65$ of $I_\mathrm{rated}$) were fixed before optimization
and are distinct from the protection thresholds (e.g.\ $I_\mathrm{trip}=3I_\mathrm{rated}$).
After the circle check the $q$-axis reference passes the 2~ms output filter and is
re-limited to $\min\!\bigl[I_{q,\max},\,\sqrt{I_\mathrm{ref,max}^{2}-(i_d^{*})^{2}}\bigr]$,
so dc-bus regulation keeps priority during deep compensation demands.
Cross-coupling is cancelled by the feedforward of equation~\eqref{eq:ff},
\begin{equation}
  v_{d,ff}=V_d-\omega L_g i_q+R_g i_d,\quad
  v_{q,ff}=V_q+\omega L_g i_d+R_g i_q,
  \label{eq:ff}
\end{equation}
and the unsaturated commands, following standard synchronous-frame design
practice~\cite{ref_ghoshal_ctrl}, are the linear PI laws of
equation~\eqref{eq:dqpipi},
\begin{equation}
  v_{d,u}=v_{d,ff}+K_{p,id}\,e_d+K_{i,id}\,\xi_d,\quad
  v_{q,u}=v_{q,ff}+K_{p,iq}\,e_q+K_{i,iq}\,\xi_q,
  \label{eq:dqpipi}
\end{equation}
with $e_d=i_d^{*}-i_d$ and $e_q=i_q^{*}-i_q$, followed by equation~\eqref{eq:vsat} and directional conditional-integration
anti-windup on the inner loops. The DQ-PI controller is retained as a practical
baseline; a residual mean reactive offset can remain under the harsher stochastic
arc, as reported in Section~\ref{sec:results}.

\subsection{Lyapunov Nonlinear Controller}
The Lyapunov controller commands active and reactive power in the outer loops
and uses a nonlinear inner current law. Its six tuned gains are
$\symbf{x}_\mathrm{Lyap}=[K_\mathrm{droop},K_{q,p},K_{q,i},k_{1,i},k_{i,i},k_{2,i}]$.
The fixed dc-link gains are $K_{v,p}=4480$ and $K_{v,i}=59700$. The raw dc error is
$e_{dc,\mathrm{raw}}=V_\mathrm{dc,ref}-V_\mathrm{dc,meas}$ (instantaneous, not
filtered). Within a half-width $V_\mathrm{taper}=0.002\,V_\mathrm{dc,ref}$ a quadratic
taper $e_{dc,\mathrm{eff}}=e_{dc,\mathrm{raw}}|e_{dc,\mathrm{raw}}|/V_\mathrm{taper}$
applies, and $e_{dc,\mathrm{eff}}=e_{dc,\mathrm{raw}}$ outside it, which reduces
small limit cycling without a hard dead-band; $\xi_v$ accumulates
$e_{dc,\mathrm{eff}}$. The active-power command uses the loss estimate of
equation~\eqref{eq:lyaploss},
\begin{equation}
  P_\mathrm{loss,ff}=P_0+K_\mathrm{loss}(i_d^2+i_q^2),\qquad
  P_0=1.5~\mathrm{kW},\quad K_\mathrm{loss}=0.002~\mathrm{W/A^2},
  \label{eq:lyaploss}
\end{equation}
and the outer active- and reactive-power commands are given by equations~\eqref{eq:pcmd} and~\eqref{eq:qcmd},
\begin{align}
  P_\mathrm{cmd}&=P_\mathrm{loss,ff}+K_{v,p}\,e_{dc,\mathrm{eff}}+K_{v,i}\,\xi_v,
   \label{eq:pcmd}\\
  Q_\mathrm{cmd}&=K_{q,p}\,e_Q+K_{q,i}\,\xi_q,\qquad
  e_Q=Q_\mathrm{ref,var}+K_\mathrm{droop}(V_\mathrm{ref}-V_\mathrm{pu})-Q_\mathrm{meas,var}.
   \label{eq:qcmd}
\end{align}
Reactive power is logged in kVAr and converted internally to var; the command is
protected by amplitude anti-windup and a $5\times10^7$~var/s slew limit. The power
commands become current references as
\begin{equation}
  i_{d,raw}^{*}=s_p\frac{P_\mathrm{cmd}}{1.5V_\mathrm{eff}},\qquad
  i_{q,raw}^{*}=s_q\frac{Q_\mathrm{cmd}}{1.5V_\mathrm{eff}},
  \quad s_p=-1,\;s_q=+1,
  \label{eq:lyapiref}
\end{equation}
where $V_\mathrm{eff}=\max(\sqrt{V_d^2+V_q^2},60~\mathrm{V})$ and the signs follow
the Park-transform and network power conventions used throughout. Both references
are slew limited to $10^5$~A/s. If their magnitude exceeds $I_\mathrm{ref,max}=486.9$~A, the
Lyapunov controller uses the radial projection of equation~\eqref{eq:lyapcircle},
\begin{equation}
  \begin{bmatrix}i_d^*\\ i_q^*\end{bmatrix}
  \leftarrow
  \frac{I_\mathrm{ref,max}}{\sqrt{(i_d^*)^2+(i_q^*)^2}}
  \begin{bmatrix}i_d^*\\ i_q^*\end{bmatrix},
  \label{eq:lyapcircle}
\end{equation}
which differs from the $d$-axis-priority circle used by DQ-PI. The protection
logic uses $I_\mathrm{trip}=3I_\mathrm{rated}\approx2086.8$~A, a 50~ms delayed
trip, an immediate hard threshold at $2I_\mathrm{trip}$, and reset below
$0.8I_\mathrm{trip}$. These limits are fixed and are not optimizer decisions.

The inner loop replaces linear PI with the smooth-sign nonlinear law of
equation~\eqref{eq:nl},
\begin{equation}
  \sigma_x=\frac{e_x}{|e_x|+\phi_i},\quad
  n_x=k_{2,i}\,\sigma_x\,(|e_x|+10^{-6})^{\alpha},\quad x\in\{d,q\},
  \label{eq:nl}
\end{equation}
with fixed boundary-layer constant $\phi_i=5$~A and fixed exponent
$\alpha=0.85<1$ (a smooth, sublinearly weighted error correction), while the
nonlinear gain $k_{2,i}$ is optimizer-tuned; this gives the control voltages of
equation~\eqref{eq:lyapv},
\begin{equation}
  v_{x,u}=v_{x,ff}+L_g\bigl(k_{1,i}\,e_x+n_x\bigr)+k_{i,i}\,\xi_{ix}.
  \label{eq:lyapv}
\end{equation}
Here $\xi_{ix}$ is the current-error integral state, defined by its time
derivative $\dot{\xi}_{ix}=e_x$ (equivalently $\xi_{ix}=\int_0^{t}e_x\,d\tau$).
The integral states $\xi_d$ and $\xi_q$ in the DQ-PI law of
equation~\eqref{eq:dqpipi} are defined the same way, with $\dot{\xi}_d=e_d$ and
$\dot{\xi}_q=e_q$.

\paragraph{Nominal stability analysis.} The unsaturated inner current-error
subsystem is analyzed with the extended storage function of
equation~\eqref{eq:V}, which retains the integral states rather than discarding
them,
\begin{equation}
  V(\symbf{e},\symbf{\xi})=\tfrac{1}{2}\bigl(e_d^{2}+e_q^{2}\bigr)
  +\frac{k_{i,i}}{2L_g}\bigl(\xi_{id}^{2}+\xi_{iq}^{2}\bigr)\ge0.
  \label{eq:V}
\end{equation}
For $k_{i,i}>0$ this $V$ is positive definite and radially unbounded in the full
error state $(\symbf{e},\symbf{\xi})$. The result established below is stated as
Proposition~1.

\medskip
\noindent\textbf{Proposition 1.} \textit{For $k_{1,i}>0$, $k_{i,i}>0$ and
$k_{2,i}>0$, the origin of the nominal disturbance-free unsaturated
current-error subsystem \eqref{eq:edyn} is globally asymptotically stable.}
\medskip The inductor dynamics are
$L_g\dot{e}_x=-v_{x,\mathrm{ctrl}}+d_x$, where $d_x$ collects the arc-driven
disturbance and the feedforward residual. Substituting the control voltage of
equation~\eqref{eq:lyapv} and dividing by $L_g$ gives the closed-loop error
dynamics of equation~\eqref{eq:edyn},
\begin{equation}
  \dot e_x=-k_{1,i}e_x-n_x-\frac{k_{i,i}}{L_g}\xi_{ix}+\tilde d_x,
  \qquad \dot\xi_{ix}=e_x,
  \label{eq:edyn}
\end{equation}
with $\tilde d_x=d_x/L_g$ the disturbance scaled by the interface inductance.
Differentiating equation~\eqref{eq:V} along equation~\eqref{eq:edyn}, the
cross-terms $-\left(k_{i,i}/L_g\right)e_x\xi_{ix}$ arising from the integral
feedback and $+\left(k_{i,i}/L_g\right)\xi_{ix}\dot\xi_{ix}$ arising from the
added storage term cancel exactly, giving equation~\eqref{eq:Vdot1},
\begin{equation}
  \dot V=-k_{1,i}\lVert\symbf{e}\rVert^{2}
         -\sum_{x\in\{d,q\}} e_x n_x
         +\symbf{e}^{\top}\tilde{\symbf{d}}.
  \label{eq:Vdot1}
\end{equation}
The integral state therefore no longer has to be suppressed to obtain a sign
conclusion. Because $n_x$ has the same sign as $e_x$, the product $-e_xn_x\le0$,
so the nonlinear term is dissipative. For the nominal disturbance-free
unsaturated subsystem ($\tilde{\symbf{d}}=\symbf{0}$), substituting the exact
form of $n_x$ from equation~\eqref{eq:nl} reduces equation~\eqref{eq:Vdot1} to
equation~\eqref{eq:Vdot},
\begin{equation}
  \dot V_{\mathrm{nom}}
   = -k_{1,i}\bigl(e_d^{2}+e_q^{2}\bigr)
     -k_{2,i}\!\sum_{x}\frac{e_x^{2}\,(|e_x|+\epsilon)^{\alpha}}{|e_x|+\phi_i}
   \le 0,
  \label{eq:Vdot}
\end{equation}
with $\epsilon=10^{-6}$~A. This is negative semidefinite in
$(\symbf{e},\symbf{\xi})$, and $\dot V_{\mathrm{nom}}=0$ if and only if
$\symbf{e}=\symbf{0}$ for positive $k_{1,i}$ and $k_{2,i}$. On any trajectory
confined to $\symbf{e}\equiv\symbf{0}$ one has $n_x=0$ and $\dot e_x=0$, so
equation~\eqref{eq:edyn} forces $\left(k_{i,i}/L_g\right)\xi_{ix}=0$ and hence
$\symbf{\xi}=\symbf{0}$ for $k_{i,i}>0$. The largest invariant set contained in
$\{\dot V_{\mathrm{nom}}=0\}$ is therefore the origin
$(\symbf{e},\symbf{\xi})=(\symbf{0},\symbf{0})$, and since $V$ is radially
unbounded, LaSalle's invariance principle~\cite{ref_khalil} gives global
asymptotic stability, which proves Proposition~1. The required gain conditions
are satisfied by both Lyapunov champions of Supplementary Table~S4.

The linear gain $k_{1,i}$ gives exponential-type decay, the nonlinear term adds a
sign-consistent dissipative contribution growing sublinearly with error, and the
integral term now contributes to the certificate rather than being set aside.
The scope is explicit. Proposition~1 establishes global asymptotic stability of
the \emph{nominal unsaturated current-error subsystem}; it does not constitute a
global stability proof for the complete EAF--grid--STATCOM system, whose
outer-loop dynamics, converter saturation, dc-link dynamics, delays and arc
disturbances all remain outside the nominal certificate. Behavior under the arc disturbance, reference
limiting and voltage saturation is assessed numerically rather than through an
input-to-state-stability bound, equation~\eqref{eq:Vdot1} showing the
disturbance as an unmatched forcing term that is not dominated analytically
here. Supplementary Section~\SMproof{} gives the intermediate algebra. An
analogous dissipativity derivation is not developed for the DQ-PI benchmark in
this study. Together with the full reactive-power PI loop, the derived nominal
property helps explain the observed response of the Lyapunov architecture.

The two controllers therefore differ in every loop. The DQ-PI scheme
uses a filtered dc-link PI loop with a quadratic converter-loss feedforward,
outputs the $d$-axis current reference directly, and acts only on the
reactive power through a split-frequency proportional--integral law ($K_{i,Q}=K_{p,Q}/T_{iQ}$);
its inner loop is a linear PI with decoupling, protected by reference slew
limits, $d$-axis-priority current-circle limiting, and conditional-integration
anti-windup. The Lyapunov scheme senses the dc link instantaneously with a
quadratic taper, commands power that is then converted to a current reference,
adds a constant-plus-$I^2$ loss feedforward, regulates the full scaled reactive
power with voltage droop through a PI law with anti-windup and slew limiting,
uses radial current-circle limiting and the nonlinear inner law above with the nominal stability analysis of
equations~\eqref{eq:V}--\eqref{eq:Vdot}, and includes a delayed/hard trip with
hysteresis for protection. The comparison is accordingly between the implemented
practical DQ-PI baseline and the proposed Lyapunov-based architecture as a
whole, not an isolated comparison of linear-PI versus nonlinear inner current
laws.

\section{Optimization-Based Gain Tuning}
Both controllers are tuned against the same scalar objective, so that the
comparison reflects controller architecture rather than a difference in what was
optimized. The objective combines a reactive-power fluctuation term, a
flicker-band voltage-envelope term and three feasibility penalties,
\begin{equation}
  T_Q=\frac{Q_\mathrm{fluc}}{Q_\mathrm{fluc}^\mathrm{ref}},\qquad
  T_V=\frac{V_\mathrm{pcc}}{V_\mathrm{pcc}^\mathrm{ref}},
  \label{eq:TQTV}
\end{equation}
augmented by graded quality and safety penalties,
\begin{equation}
  J_\mathrm{perf}=0.60\,T_Q+0.40\,T_V,\qquad
  J_\mathrm{tot}=J_\mathrm{perf}+P_{Q}+P_{V_\mathrm{dc}}+P_\mathrm{safety}.
  \label{eq:J}
\end{equation}
where $T_Q$ and $T_V$ are normalized fluctuation measures evaluated over a
common short optimization window and $P_Q$, $P_{V_\mathrm{dc}}$ and
$P_\mathrm{safety}$ penalize reactive-power tracking error, dc-link excursion
and converter safety-limit violation. The exact feature definitions, the
flicker-band filters, the moving-window construction and the individual penalty
formulae are given in Supplementary Section~\SMobj, together with the optimizer
mechanics, the search bounds and the seed protocol.

\begin{figure}[!htbp]
\centering
\includegraphics[width=\linewidth]{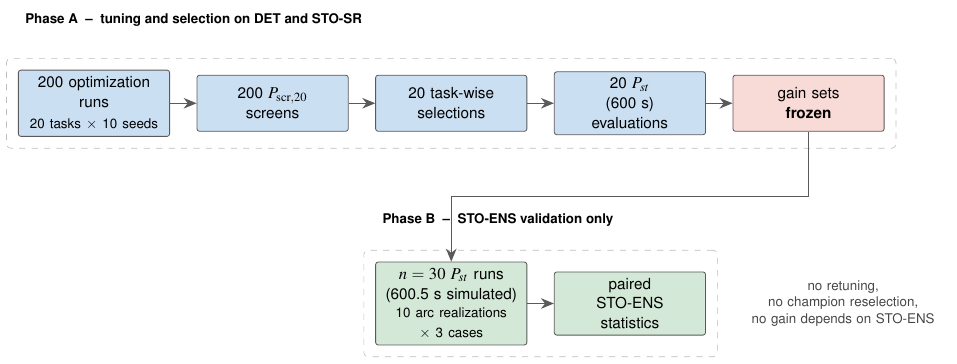}
\caption{Two-phase evaluation pipeline. Phase~A performs all tuning and
selection on the deterministic EAF (DET) and the fixed-seed stochastic-mode
realization (STO-SR), and ends with the selected gain sets frozen. Phase~B is
validation only: the frozen gain sets are re-evaluated on the stochastic
ensemble (STO-ENS) over ten arc realizations $\times$ three cases
(uncompensated, DQ-PI, Lyapunov), summarized in Table~\ref{tab:corrected}. No gain was retuned and no champion was
reselected against STO-ENS, so no reported gain set depends on it. The
$600.5$~s in the Phase~B box is the archived stop time, of which the first
$0.5$~s is start-up, leaving the same $600$~s IEC window as Phase~A. The
search objective $J_\mathrm{tot}$, the 20~s screening index
\pscr{}, and the 600~s IEC index \pst{} are distinct quantities; only the 20
task-wise selected gain sets are simulated for 600~s in Phase~A.}
\label{fig:pipeline}
\end{figure}

The frozen campaign used PSO~\cite{ref_pso}, GWO~\cite{ref_gwo},
FP-SHADE~\cite{ref_lshade}, EMBGO~\cite{ref_embgo} and REMBGO~\cite{ref_rembgo}
under a common protocol. Every optimizer receives an identical budget of $1040$
objective evaluations, a common warm start and ten matched seeds per
configuration, under both arc modes
and for both controller architectures. The resulting population of gain sets
then passes through a frozen three-stage selection pipeline: all optimized sets
are screened with a $20$~s flicker index, one task-wise champion is selected per
optimizer--controller--arc combination, and each selected set is evaluated over
$600$~s with the IEC~61000-4-15 flickermeter of
Section~\ref{sec:flicker}. Selection was completed and frozen before any
ensemble result existed, and no gain set was retuned or reselected
afterwards. Fig.~\ref{fig:pipeline} summarizes the pipeline.

\section{Long-Duration Flickermeter Evaluation}\label{sec:flicker}
\label{sec:val}
Flicker severity is evaluated with a digital IEC~61000-4-15 flickermeter
signal-processing chain embedded in the simulation model. The PCC phase voltage
reaches the flickermeter through a unity-gain measurement path, and the
flickermeter is configured with a nominal phase voltage
$V_{\mathrm{nom,ph}}=11\,\mathrm{kV}/\sqrt{3}=6.351$~kV, a 50~Hz nominal
frequency and a $10^{-4}$~s sampling interval, with the normalization required
by the standard performed internally. The implementation
demodulates the PCC voltage, applies the lamp--eye--brain weighting and squaring
stages, and forms the instantaneous flicker sensation $S(t)$; the statistical
block sorts $S(t)$ into a cumulative probability function (CPF) and reads the
short-term index from the standard percentile combination
($P_{0.1}$, $P_{1s}$, $P_{3s}$, $P_{10s}$, $P_{50s}$). The saved flicker-sensation records reproduce the exported long-duration
\pst{} values to numerical precision ($1.812156$ versus $1.812160$
uncompensated, $0.634559$ versus $\approx0.6346$ for the selected EMBGO-tuned
Lyapunov case), so the deterministic EAF values retained in
Table~\ref{tab:overview} are $\pst{}=1.8122$ uncompensated and $0.6346$
compensated. Analyzer displays and the $<0.001$ display-to-export difference are
given in Supplementary Fig.~S1 and Section~\SMtime.

Two measurement roles are kept strictly separate. All 200 optimized gain sets
are first screened with a 20~s flicker screening index \pscr{}. This index
applies the same IEC flickermeter processing chain over a shortened 20~s
observation interval; it is a comparative screening quantity, used only to rank
candidate gain sets, and is not reported as an IEC-compliant 10-min \pst{}. For
each of the 20 tasks, the gain set with the lowest \pscr{} is selected. The
headline \pst{} values reported in this paper are then obtained from full 600~s
EMT simulations of the 20 task-wise selected gain sets, corresponding to the standard IEC
short-term evaluation interval. Controller-level screening variability across the
50 pooled optimizer--seed runs is summarized in Supplementary Table~S6. The
value $\pst{}=1.0$ is used as a commonly adopted perceptibility benchmark; no
formal network-compliance claim is made.

Supplementary Table~S5 shows the cumulative window estimate converging to the
600~s value: the compensated champions are already close to it within the first
few windows while the uncompensated index rises gently, which is why the 600~s
value and not the 20~s screening value carries every headline comparison.
Plant-model parity was verified by running the uncompensated system in both
controller test benches, giving numerically identical $S(t)$ and \pst{}
(difference $<10^{-13}$), so compensated differences arise from the controller
architectures rather than plant mismatch.

Table~\ref{tab:overview} lists the headline champion for each controller--arc
combination. Each optimizer task first contributes its lowest-\pscr{} seed to the
600-s evaluation set; the headline champion is then the vector with the lowest
600~s \pst{} among those five task-wise selections. The champion gain vectors
appear in Supplementary Table~S4; the optimizer column names only the source of
each selected vector. An optimizer-level statistical comparison on the search
objective and the 20~s screening endpoint is outside the scope of this paper.

Within the STO-SR single-trajectory evaluation the result appeared
regime-dependent. Under the deterministic arc the Lyapunov champion
reaches $\pst{}=0.635$ ($65.0\%$ below the uncompensated $1.812$), against the
DQ-PI champion at $0.680$ ($62.5\%$); under STO-SR the ordering
appeared to reverse, the DQ-PI champion reaching $\pst{}=0.800$ (a $64.6\%$
reduction from the uncompensated $2.262$), below the Lyapunov champion at $0.889$
($60.7\%$). All twenty task-wise selected gain sets yield $\pst{}<1.0$ in their 600~s
simulations, with the worst selected case at $0.930$. The controller-level
ordering is consistent with the 20~s screening statistics of
Supplementary Table~S6, which pool the five optimizers and ten seeds ($n=50$)
per controller--EAF combination; those statistics characterize variability across
the pooled optimizer--seed runs at the screening window and are not multi-seed
600~s measurements.

\begin{table}[!htbp]
\centering
\caption{Long-duration 600~s \pst{} evaluation of the task-wise selected gain sets, with
the reduction relative to the uncompensated system. The optimizer column
identifies only the source of each selected gain set; an optimizer-level
comparison on the search objective and the 20~s screening endpoint is outside
the scope of this paper. Bold marks the lower optimized \pst{} within each arc condition. DET is the
deterministic EAF and STO-SR the fixed-seed stochastic-mode realization
(Table~\ref{tab:eafcases}); the stochastic ensemble STO-ENS was not used for
tuning and appears in Table~\ref{tab:corrected}.}
\label{tab:overview}
\setlength\tabcolsep{5pt}\renewcommand{\arraystretch}{1.1}\small
\begin{tabular}{lcccc}
\toprule
\textbf{Case} & \textbf{Uncomp.} & \textbf{Optimized} & \textbf{Reduction} & \textbf{Optimizer}\\
\midrule
DET, DQ-PI        & 1.812 & 0.680 & 62.5\% & GWO\\
DET, Lyapunov     & 1.812 & \textbf{0.635} & \textbf{65.0\%} & EMBGO\\
STO-SR, DQ-PI     & 2.262 & \textbf{0.800} & \textbf{64.6\%} & FP-SHADE\\
STO-SR, Lyapunov  & 2.262 & 0.889 & 60.7\% & EMBGO\\
\bottomrule
\end{tabular}
\end{table}

Each row of Table~\ref{tab:overview} is a separately tuned gain vector: the four
cases are independent optimizer--condition champions, selected task-wise, so the
DET and STO-SR entries do not share a common gain set and are not two evaluations
of one controller tuning. Table~\ref{tab:corrected} is therefore not a third
column of Table~\ref{tab:overview}: it carries forward only the two STO-SR
champions, frozen and without retuning, onto the STO-ENS realizations.

Within the five STO-SR optimizer-task champions, this trajectory-specific
ordering was not attributable solely to the single best champion.
Supplementary Table~S7 shows that, averaged over all five optimizer-task
champions of each combination, the Lyapunov mean is lower under the deterministic
arc (0.647 versus 0.701) and the DQ-PI mean is lower under STO-SR
arc (0.813 versus 0.906). Section~\ref{sec:corrected} subsequently tests whether
that the STO-SR ordering generalizes to the STO-ENS realizations. This is a descriptive controller-level summary; an
inferential optimizer comparison, on the search objective and the 20~s screening
endpoint, is outside the scope of this paper.

\FloatBarrier
\subsection{Independent Stochastic-Ensemble Validation}
\label{sec:corrected}
The single-trajectory stochastic comparison above rests on one arc realization,
which the generator audit of Section~\ref{sec:eaf} showed to be periodic. To test
whether the reported controller ordering generalizes beyond that trajectory, the
two selected controller vectors were frozen exactly as published --- the
FP-SHADE-tuned DQ-PI set (task~10, optimizer seed~5) and the EMBGO-tuned Lyapunov
set (task~16, optimizer seed~5) --- and re-evaluated on ten pre-specified
independent stochastic arc realizations. No gain was retuned, no champion was
reselected, and the arc bases were fixed before any corrected result was
computed. For every realization the matching uncompensated case was also
simulated, giving a paired ensemble of thirty $600.5$~s runs. Flicker severity
was obtained from the same IEC~61000-4-15 statistical block applied to the
retained instantaneous flicker-sensation record over $[0.5,600.5]$~s.

Table~\ref{tab:corrected} summarizes the ensemble. The corrected uncompensated
system is highly repeatable, with $\pst{}=2.2225\pm0.0021$, close to the STO-SR
value of $2.2623$ ($-1.76\%$). The frozen DQ-PI controller gives
$\pst{}=0.8491\pm0.0079$ and the frozen Lyapunov controller
$\pst{}=0.8527\pm0.0024$, corresponding to mean reductions of $61.79\%$ and
$61.63\%$ relative to the STO-ENS uncompensated baseline. Neither controller
saturated in any of the thirty runs.

\begin{table}[!htbp]
\centering
\caption{Stochastic ensemble (STO-ENS) over ten pre-specified arc realizations
(600.5~s each, frozen gain sets, no retuning, no champion reselection). The
STO-SR column is the archived fixed-seed stochastic-mode value for the same gain
set, shown for reference only.}
\label{tab:corrected}
\setlength\tabcolsep{4.5pt}\renewcommand{\arraystretch}{1.08}\small
\begin{tabular}{lcccccc}
\toprule
\textbf{Case} & \textbf{STO-SR} & \textbf{STO-ENS} & \textbf{SD} &
\textbf{95\% CI} & \textbf{Range} & \textbf{Shift}\\
 & \pst{} & mean \pst{} & & & & vs STO-SR\\
\midrule
Uncompensated      & 2.2623 & 2.2225 & 0.0021 & 2.2209--2.2240 & 2.2195--2.2253 & $-1.76\%$\\
DQ-PI (T10)        & 0.7999 & 0.8491 & 0.0079 & 0.8434--0.8548 & 0.8398--0.8677 & $+6.15\%$\\
Lyapunov (T16)     & 0.8892 & 0.8527 & 0.0024 & 0.8509--0.8544 & 0.8465--0.8551 & $-4.11\%$\\
\midrule
DQ-PI reduction    & 64.64\% & 61.79\% & -- & -- & 60.92--62.24\% & --\\
Lyapunov reduction & 60.70\% & 61.63\% & -- & -- & 61.50--61.91\% & --\\
\bottomrule
\end{tabular}
\end{table}

The ensemble changes the interpretation of the single-trajectory result. On the
STO-SR realization the DQ-PI champion was $10.04\%$ below the Lyapunov champion
in \pst{}. Over STO-ENS the mean paired difference
$\Delta\pst{}=\pst{}_{,\mathrm{DQPI}}-\pst{}_{,\mathrm{Lyap}}$ falls to
$-0.003527$, which is $0.41\%$ of the Lyapunov mean. DQ-PI gives the lower \pst{}
on seven of the ten realizations, but the paired difference is not statistically
detectable ($t=-1.323$, $p=0.218$, $n=10$; Wilcoxon signed-rank $p=0.193$;
Cohen's $d_z=-0.42$), and the $95\%$ confidence interval $[-0.00956,+0.00250]$
spans zero. The STO-ENS experiment therefore provides no evidence of a
systematic difference in mean 600-s flicker severity between the two frozen
controllers. This is a statement about the absence of a detectable difference in
this ten-realization ensemble; it is not a formal equivalence claim, which would
require a pre-specified equivalence margin and a corresponding test.

\begin{figure}[!htbp]
\centering
\includegraphics[width=\linewidth]{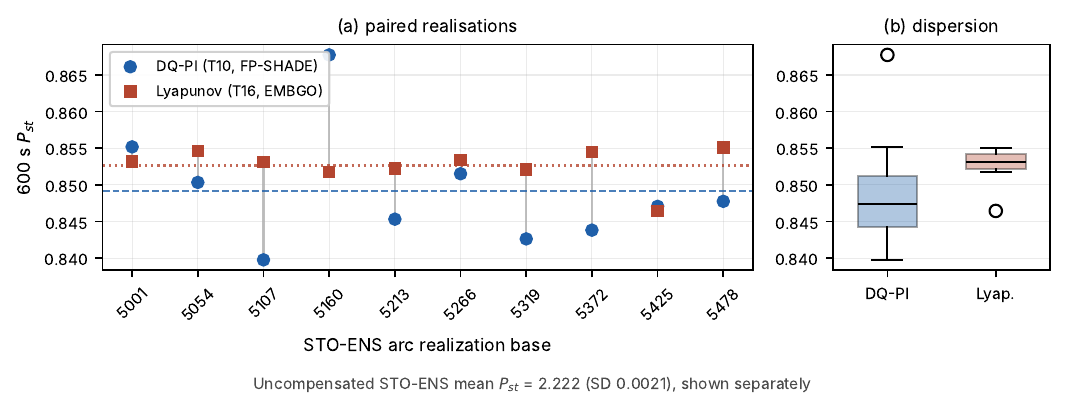}
\caption{Stochastic ensemble. (a) Paired 600~s \pst{} for the
frozen DQ-PI and Lyapunov gain sets on each of the ten pre-specified arc
realizations; grey segments join the paired values and the dashed and dotted
lines mark the two ensemble means. (b) Distribution of the same values; the
open circles are the realizations beyond the $1.5$ interquartile-range
whiskers, which are also plotted in (a) and are retained in every reported
statistic. The STO-ENS uncompensated
baseline is reported separately in Table~\ref{tab:corrected} and lies off the
plotted scale.}
\label{fig:corrected}
\end{figure}

A separate distinction does emerge in dispersion across realizations. The
standard deviation of the DQ-PI \pst{} is $0.00794$ against $0.00245$ for
Lyapunov. The descriptive quantities are a standard-deviation ratio of $3.245$
and a variance ratio of $0.095$, so the DQ-PI variance is about $10.5$ times
larger. A post-hoc paired Pitman--Morgan test on the ten realizations yielded
$r=0.827$, $t=4.16$, $p=0.0032$. This dispersion analysis was not pre-specified in
the validation plan; with $n=10$ and a post-hoc hypothesis, the descriptive ratios
rather than the $p$-value should be read as the headline, and the result is
reported as a secondary, exploratory observation that motivates, but does not
substitute for, a pre-specified dispersion study.
Fig.~\ref{fig:corrected} shows the paired values and the two distributions.

The asymmetry of the change is itself informative. Relative to STO-SR
trajectory the STO-ENS \pst{} rises by $6.15\%$ for DQ-PI, falls by
$4.11\%$ for Lyapunov, and falls by only $1.76\%$ for the uncompensated system.
The collapse of the controller gap is therefore not attributable to a uniform
change in arc severity: the two closed loops respond differently to the STO-ENS
perturbation, which is the substantive reason a single stochastic trajectory is
an unreliable basis for a robustness ordering.

The 20~s STO-ENS screening runs provide a consistency check and are used only
as such. Across the five frozen task winners of each controller, the DQ-PI group
gives a mean screening index of $0.8424$ and the Lyapunov group $0.8744$, but
dispersion depends strongly on the individual gain vector: the DQ-PI standard
deviations range from $0.0132$ (task~14) to $0.0585$ (task~18), while the
Lyapunov winners are more tightly clustered. These short-window results are
reported as sensitivity evidence only. The 600~s comparison retains the
historically selected task-10 DQ-PI and task-16 Lyapunov vectors, and no
controller was reselected after the corrected results were seen.

\FloatBarrier
\section{Time-Domain Results}
\label{sec:results}
The 600~s IEC flickermeter value is the primary performance indicator; the
time-domain records are supporting diagnostics that explain how each controller
reaches its long-duration flicker result. Every metric in this section is
computed over the same 600~s evaluation run that produced the \pst{} value of the
corresponding champion, so signal metrics and \pst{} always share one gain set.
The metrics are summarized in Supplementary Table~S2.

The two controllers differ most in the dc-link and mean-reactive-power
quantities. The Lyapunov champions keep the fluctuating source reactive power to
1.18 and 1.70~kVAr and the mean source reactive power close to zero
($|\overline Q|<0.02$~kVAr) in both arc modes, and hold the maximum dc-link
voltage deviation to 0.205\% and 0.266\%. The DQ-PI champions keep a comparable
fluctuating reactive power (2.23 and 2.04~kVAr) but retain a mean offset of 4.3
and 25.8~kVAr, and their dc-link voltage deviation is larger, 4.98\% and 5.98\%.
No voltage-vector saturation occurs in any compensated case.

The comparison is balanced by a tracking trade-off. The DQ-PI champions achieve
lower normalized current-tracking IAE (0.0339 and 0.0352) than the Lyapunov
champions (0.0438 and 0.0569) in both arc modes, because the linear inner PI
tracks the current reference tightly. The Lyapunov champions instead give lower
voltage THD (1.63\% and 0.889\%, versus 1.80\% and 1.93\% for DQ-PI). Lower
inner-current IAE therefore does not by itself yield the lowest \pst{}:
current-tracking accuracy alone is insufficient, and the long-duration flicker
reflects the combined outer-loop reactive-power regulation, dc-link dynamics, and
frequency-selective PCC-voltage response.

\subsection{PCC Voltage and Source Reactive Power}
\begin{figure*}[!htbp]
\centering
\begin{minipage}{0.49\linewidth}\centering
  \includegraphics[width=\linewidth]{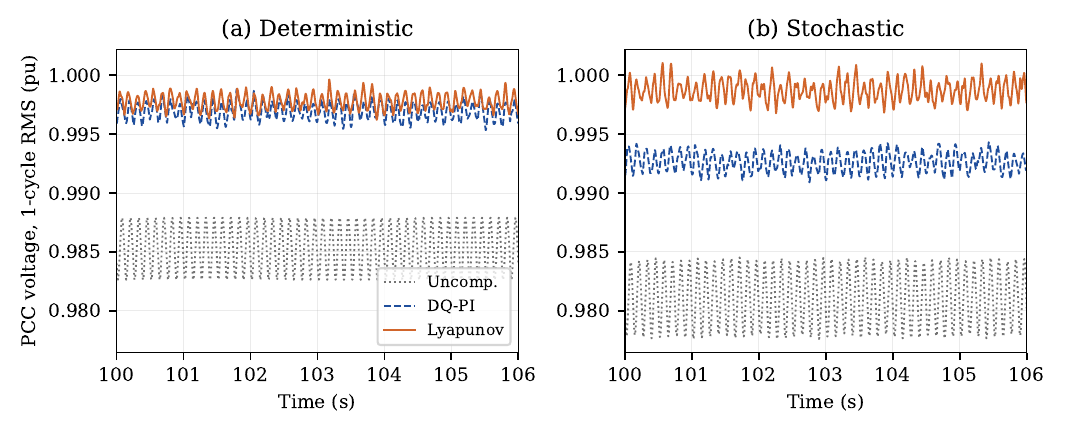}\\[2pt]{\footnotesize (a) PCC voltage envelope}
\end{minipage}\hfill
\begin{minipage}{0.49\linewidth}\centering
  \includegraphics[width=\linewidth]{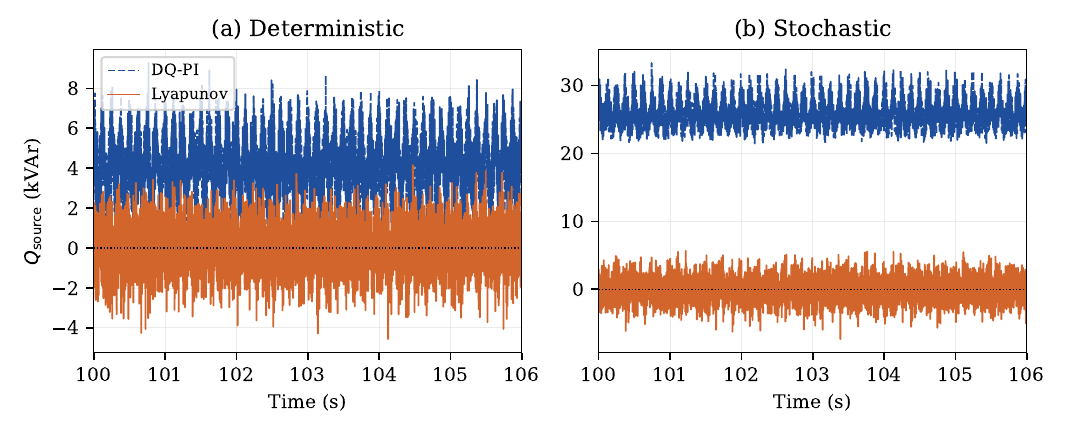}\\[2pt]{\footnotesize (b) source reactive power}
\end{minipage}

\vspace{4pt}
\begin{minipage}{0.49\linewidth}\centering
  \includegraphics[width=\linewidth]{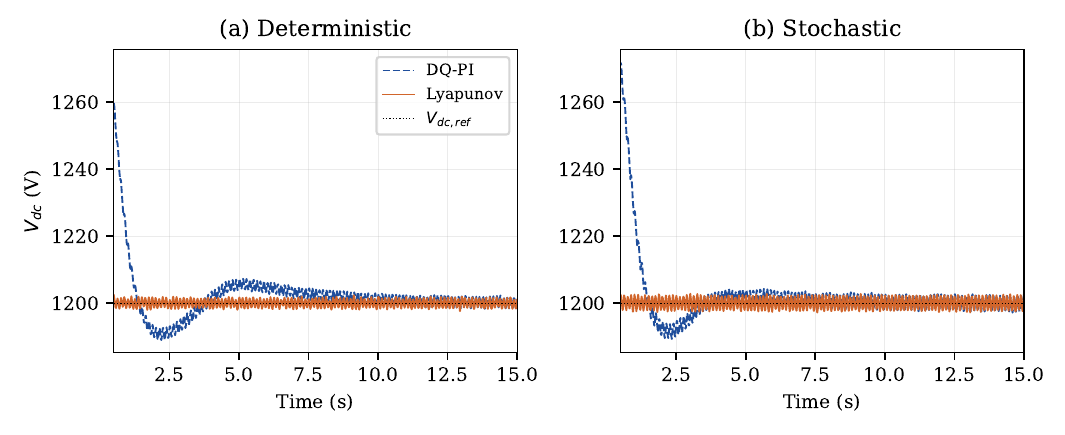}\\[2pt]{\footnotesize (c) dc-link voltage}
\end{minipage}\hfill
\begin{minipage}{0.49\linewidth}\centering
  \includegraphics[width=\linewidth]{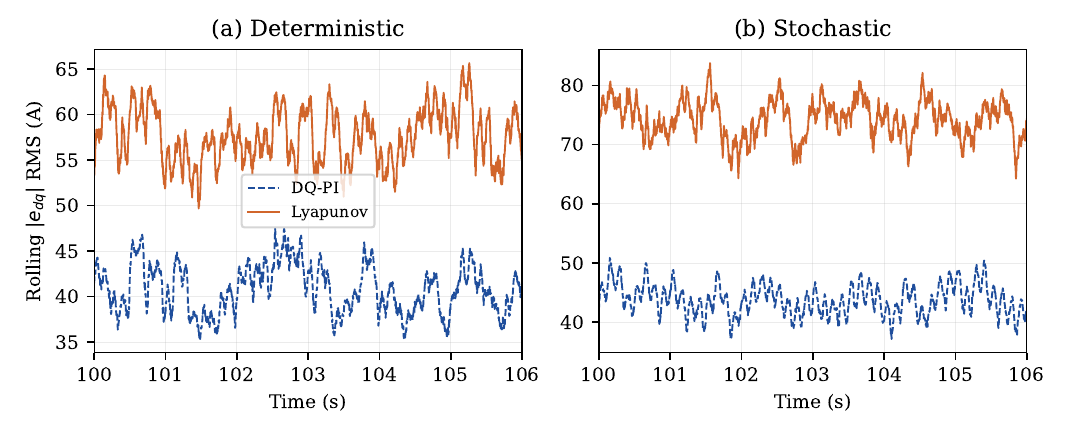}\\[2pt]{\footnotesize (d) current-tracking error}
\end{minipage}
\caption{Closed-loop response of the two controllers under deterministic and
stochastic arc conditions: (a) PCC voltage envelope, (b) source-side reactive
power, (c) dc-link voltage, (d) inner current-tracking error. The Lyapunov design
holds source reactive power and dc-link voltage tighter; the DQ-PI design tracks
the inner current more accurately.}
\label{fig:fourpanel}
\end{figure*}

Fig.~\ref{fig:fourpanel} compares the one-cycle RMS PCC voltage for the
uncompensated and optimized cases. Both compensated controllers raise the mean
PCC voltage toward 1~pu: the Lyapunov champions reach 0.9977 and 0.9987~pu and
the DQ-PI champions 0.9970 and 0.9926~pu, against uncompensated values of 0.9853
and 0.9811~pu. These RMS statistics are not numerically equivalent to \pst{},
because the IEC flickermeter applies frequency-selective lamp--eye--brain
weighting and statistical processing; they identify only the physical reduction
in PCC voltage modulation that accompanies the long-duration flicker result.

Fig.~\ref{fig:fourpanel} makes the structural difference explicit. The
Lyapunov full reactive-power PI regulates both the mean and the fluctuating
components and drives the mean source reactive power to near zero in both arc
modes. The split-frequency DQ-PI loop suppresses the fluctuating component to a
similar level but leaves a mean offset that is small in the deterministic case
(4.3~kVAr) and larger under the stochastic arc (25.8~kVAr), where the integral
path is constrained by its reactive-current headroom and slow-error deadband.
This standing offset is the main structural difference between the two reactive
loops and is accompanied by the larger dc-link deviation of the DQ-PI champions. The source-reactive-power traces in Fig.~\ref{fig:fourpanel}(b), together with the
rolling RMS values in Supplementary Fig.~S2, show this residual directly.

The three-phase PCC voltages remain balanced in all investigated cases
(voltage unbalance factor below 0.25\%). The compensated cases exhibit a modest
increase in voltage distortion due to converter harmonic injection; the mean
three-phase voltage THD of the champions stays within 0.89\%--1.93\%, as reported
in Supplementary Table~S2.

\subsection{DC-Link Regulation and Current Tracking}
Fig.~\ref{fig:fourpanel} shows the dc-link voltage responses. The Lyapunov
champions stay close to the 1200~V reference, with startup-inclusive maximum
deviations of 2.5~V (0.205\%) and 3.2~V (0.266\%) under deterministic and
stochastic operation. The DQ-PI champions show larger startup-inclusive
excursions, reaching 59.8~V (4.98\%) and 71.7~V (5.98\%). The main difference in
Fig.~\ref{fig:fourpanel} occurs during the initial charging and settling
interval; after this transient, both architectures maintain the dc link close to
the 1200~V reference. The reported 4.98\% and 5.98\% DQ-PI values should
therefore be interpreted as startup-inclusive maximum deviations rather than
steady-state regulation errors, and it is these startup excursions that cause the
DQ-PI champions to exceed the $3\%$ free dc-link region and incur the graded
$P_{V_\mathrm{dc}}$ penalty during optimization, while remaining below the $8\%$ hard
feasibility limit. Lyapunov nonetheless has the smaller startup-inclusive maximum
deviation.

Fig.~\ref{fig:fourpanel}(d) presents the rolling RMS magnitude of the
combined \(dq\)-axis current-tracking error, while Supplementary Fig.~S3
compares the reference and measured \(d\)- and \(q\)-axis currents.
The DQ-PI champions track the current reference more accurately in both arc
modes. The \(q\)-axis current mainly supplies the reactive-power compensation
demand and the \(d\)-axis current supports dc-link energy regulation. The
Lyapunov champions show a larger internal tracking error but a smaller
startup-inclusive dc-link excursion and near-zero mean source reactive power; the tracking-error result is
therefore reported as a trade-off rather than as evidence of overall superiority
of either controller.

\subsection{Integrated Interpretation and Study Limitations}
The long-duration and time-domain results support a regime-dependent
interpretation, but the two regimes are now supported by different amounts of
evidence. Under the deterministic arc the Lyapunov architecture gives the lower
600~s \pst{} (0.635 versus 0.680 for DQ-PI); that arc mode is bit-identical
between the two sampler implementations, so this ordering is
unaffected by the generator audit and stands as reported. Under the stochastic
arc, the single STO-SR realization gave the lower long-duration \pst{} to the
split-frequency DQ-PI design (0.800 versus 0.889), but the STO-ENS ten-realization
ensemble of Section~\ref{sec:corrected} does not sustain that ordering as a
detectable mean effect: the gap narrows to $0.41\%$ of the Lyapunov mean with
$p=0.218$.

The physical reasoning behind each architecture is unchanged: the deterministic
Lyapunov advantage is consistent with its full source-reactive-power regulation,
which removes the standing bias and most of the fluctuating component, while the
DQ-PI split-frequency loop trades a little of that for lower inner-current error.
What the ensemble changes is the inference drawn from a single trajectory --- the
two mechanisms deliver near-identical mean mitigation, and the measurable
difference lies in realization sensitivity rather than mean severity. Optimizer-to-optimizer
variability in the search objective and the 20~s screening endpoint is a separate
question from the 600~s champion comparison reported here.

The trade-offs are clear: DQ-PI gives lower inner-current IAE in both arc modes,
Lyapunov lower voltage THD and near-zero mean source reactive power. Since
$J_\mathrm{tot}$ is dominated by $T_Q$ and $T_V$ while THD and IAE are
unoptimized diagnostics, a gain set minimizing $J_\mathrm{tot}$ need not minimize
every individual metric.

The $60.7$--$65.0\%$ long-duration reductions are of the same order as earlier
converter-based EAF compensation studies; Larsson and Poumar\`ede reported about
$63\%$ reduction of their comparative indicator under random arc-length
modulation~\cite{ref_larsson}. Differences in furnace model, compensator rating,
controller structure and flicker-evaluation procedure make this a check on
physical credibility rather than a like-for-like compliance match.

The stochastic EAF remains the most demanding condition. All twenty task-wise
selected gain sets lay below the $\pst{}=1.0$ perceptibility benchmark on the
stochastic-mode realization (STO-SR) (best compensated value $0.800$), and all twenty compensated
runs of the ten-realization ensemble of Section~\ref{sec:corrected} also remain
below it, with a maximum of $0.868$. These are simulation results under the investigated conditions and are
not a formal network-compliance statement. The study remains simulation-based,
and the stochastic-ensemble validation is limited to ten pre-specified
long-duration arc realizations drawn from one EAF model at one plant operating
condition; although that ensemble supports paired inference on the two tested
controller vectors, it does not establish performance across different furnace
ratings, arc-model parameterizations, grid strengths, or measured industrial
trajectories. The
Cassie--Mayr parameters are selected to provide a representative and reproducible
stochastic operating condition rather than identified from measurements of a
specific industrial furnace, and the $1.0$ reference is a perceptibility
benchmark, not a compliance limit. Future work should extend the STO-ENS set
across operating points and furnace parameterizations, add
pre-specified dispersion testing, and validate the findings experimentally or in
hardware-in-the-loop. A post-optimization bound audit
shows measurable boundary pressure: several coordinates of the deterministic
DQ-PI champion ($K_{p,id}$, $K_{i,id}$, $K_{i,iq}$, $K_{p,Q}$ and $T_{iQ}$) lie on
their prescribed search bounds; the deterministic Lyapunov champion places
$K_\mathrm{droop}$ and $k_{i,i}$ at their lower bounds; and the reactive-integral
gain $K_{q,i}$ reaches its upper bound for the STO-SR Lyapunov champion, as it
does for most STO-SR Lyapunov seed winners. The search regions were themselves defined
around previously validated stable operating points with added margin. The
reported optima should therefore be interpreted as the best solutions obtained
within the prespecified validated gain regions rather than as unconstrained
controller optima; a wider region might yield further improvement and is left to
future work.

\FloatBarrier
\section{Conclusion}
This paper compared a split-frequency DQ-PI STATCOM controller with a
Lyapunov-based nonlinear controller for EAF flicker mitigation on an 11~kV
feeder, with gains obtained from a frozen metaheuristic campaign and long-window
severity measured by an IEC~61000-4-15 flickermeter. Retaining the integral state
in the storage function yields global asymptotic stability of the nominal
unsaturated current-error subsystem, so the integral action now contributes to
the stability certificate rather than being set aside.

Under the deterministic arc the Lyapunov controller gave the lower 600~s severity
($\pst{}=0.635$ against $0.680$). STO-SR suggested the
opposite ordering, but a post-campaign audit showed that realization to be
periodic. The two selected gain sets were therefore frozen and re-evaluated,
without retuning or reselection, on ten pre-specified STO-ENS realizations
with paired uncompensated baselines. Both controllers retained strong
mitigation, reducing the mean 600-s severity by $61.79\%$ (DQ-PI) and $61.63\%$
(Lyapunov), and neither saturated. The single-trajectory advantage did not
generalize: the mean paired difference was $-0.0035$ with a confidence interval
spanning zero and no statistically detectable difference between the
controllers. They differed instead in realization sensitivity, the Lyapunov
standard deviation across arcs being roughly one third that of DQ-PI, an
exploratory observation that was not pre-specified.

Controller conclusions drawn from a single stochastic EAF trajectory can
therefore be misleading, and ensemble-based validation is necessary when
assessing flicker-mitigation robustness. Future work should extend the STO-ENS set
across operating points and furnace parameterizations, add
pre-specified dispersion testing, and validate experimentally or in
hardware-in-the-loop.

\paragraph{Data availability.}
The numerical results and methodological details supporting the reported findings are provided in the article and supplementary material.


\begin{thebibliography}{99}
\footnotesize
\bibitem{ref_khalil} H.~K. Khalil, \emph{Nonlinear Systems}, 3rd~ed. Upper
Saddle River, NJ, USA: Prentice Hall, 2002.
\bibitem{ref_iec415} IEC (2010) IEC 61000-4-15:2010: Electromagnetic compatibility (EMC) --- Part 4-15: Testing and measurement techniques --- Flickermeter --- functional and design specifications, 2nd edn. IEC, Geneva
\bibitem{ref_virtanen} Virtanen A, Tuusa H, Aho J (2013) Performance analysis of conventional STATCOMs and STATCOMs with energy storage in electric arc furnace applications. In: Proc 28th Annual IEEE Applied Power Electronics Conference and Exposition (APEC), Long Beach, CA, USA, pp 1623--1629. https://doi.org/10.1109/APEC.2013.6520514
\bibitem{ref_orcajo} Alonso Orcajo G, et al (2020) Enhancement of power quality in an actual hot rolling mill plant through a STATCOM. IEEE Trans Ind Appl 56(3):3238--3249. https://doi.org/10.1109/TIA.2020.2976034
\bibitem{ref_hussen} Hussen M, Rajaram T (2025) A Comprehensive Review of Voltage Source Converters-Based FACTS Controllers in Hybrid Microgrids. IEEE Access 13:62961--62999. https://doi.org/10.1109/ACCESS.2025.3557961
\bibitem{ref_larsson} Larsson T, Poumar\`ede C (1999) STATCOM, an efficient means for flicker mitigation. In: Proc IEEE Power Engineering Society Winter Meeting, New York, NY, USA, vol 2, pp 1208--1213. https://doi.org/10.1109/PESW.1999.747380
\bibitem{ref_garcia} Garc\'ia-Cerrada A, Garc\'ia-Gonz\'alez P, Collantes R, G\'omez T, Anzola J (2000) Comparison of thyristor-controlled reactors and voltage-source inverters for compensation of flicker caused by arc furnaces. IEEE Trans Power Deliv 15(4):1225--1231. https://doi.org/10.1109/61.891507
\bibitem{ref_yazdani} Yazdani A, Crow ML, Guo J (2009) An improved nonlinear STATCOM control for electric arc furnace voltage flicker mitigation. IEEE Trans Power Deliv 24(4):2284--2290. https://doi.org/10.1109/TPWRD.2009.2027508
\bibitem{ref_morati} Morati M, Girod D, Terrien F, Peron V, Poure P, Saadate S (2016) Industrial 100-MVA EAF voltage flicker mitigation using VSC-based STATCOM with improved performance. IEEE Trans Power Deliv 31(6):2494--2501. https://doi.org/10.1109/TPWRD.2015.2508498
\bibitem{ref_feola} Feola L, Langella R, Papi\v{c} I, Testa A (2018) Selective interharmonic compensation to improve STATCOM performance for light flicker mitigation. IEEE Trans Power Deliv 33(5):2442--2451. https://doi.org/10.1109/TPWRD.2018.2810333
\bibitem{ref_cernan} \v{C}er\v{n}an M, M\"uller Z, Tlust\'y J, Valouch V (2021) An improved SVC control for electric arc furnace voltage flicker mitigation. Int J Electr Power Energy Syst 129:106831. https://doi.org/10.1016/j.ijepes.2021.106831
\bibitem{ref_liu_spea} Liu K, Wang F, Xu C (2025) An online SPEA-II genetic algorithm for cascaded STATCOM suppressing voltage flicker in electric arc furnace system. Electr Power Syst Res 246:111667. https://doi.org/10.1016/j.epsr.2025.111667
\bibitem{ref_kamel} Kamel OM, Diab AAZ, Do TD, Mossa MA (2020) A novel hybrid ant colony--particle swarm optimization techniques based tuning STATCOM for grid code compliance. IEEE Access 8:41566--41587. https://doi.org/10.1109/ACCESS.2020.2976828
\bibitem{ref_zhang} Zhang T, Xu X, Li Z, Abu-Siada A, Guo Y (2020) Optimum location and parameter setting of STATCOM based on improved differential evolution harmony search algorithm. IEEE Access 8:87810--87819. https://doi.org/10.1109/ACCESS.2020.2993066
\bibitem{ref_makki} Makki Abadi M, Tang H, Rashidi MM (2024) A review of simulation and numerical modeling of electric arc furnace (EAF) and its processes. Heliyon 10(11):e32157. https://doi.org/10.1016/j.heliyon.2024.e32157
\bibitem{ref_teklic} Tekli\'c AT, Filipovi\'c-Gr\v{c}i\'c B, Pavi\'c I (2017) Modelling of three-phase electric arc furnace for estimation of voltage flicker in power transmission network. Electr Power Syst Res 146:218--227. https://doi.org/10.1016/j.epsr.2017.01.037
\bibitem{ref_babaei} Babaei Z, Samet H, Jalil M (2023) An innovative approach considering active power and harmonics for modeling the electric arc furnace along with analyzing time-varying coefficients based on ARMA models. Int J Electr Power Energy Syst 153:109377. https://doi.org/10.1016/j.ijepes.2023.109377
\bibitem{ref_tlili} Tlili F, Ben Slimene M, Khlifi MA, Bacha F (2024) Study of the flicker phenomenon induced in the electric arc furnace of the steel factory in Tunisia. IEEE Access 12:174850--174858. https://doi.org/10.1109/ACCESS.2024.3502428
\bibitem{ref_baros} Baros S, Ili\'c MD (2015) Multi-objective Lyapunov-based control of a STATCOM/BESS. In: Proc 2015 IEEE Power \& Energy Society Innovative Smart Grid Technologies Conference (ISGT), Washington, DC, USA, pp 1--5. https://doi.org/10.1109/ISGT.2015.7131777
\bibitem{ref_jaipra} Jaipradidtham C, Panyapnonwirot S (2014) Stability analysis for control strategy of a STATCOM and TCSC control of nonlinear co-ordinated excitation with aperiodically sampled functional by an augmented Lyapunov--Krasovskii method. In: Proc 2014 IEEE PES Asia-Pacific Power and Energy Engineering Conference (APPEEC), Hong Kong, China, pp 1--6. https://doi.org/10.1109/APPEEC.2014.7066131
\bibitem{ref_jin} Jin Y, et al (2022) A dual-layer back-stepping control method for Lyapunov stability in modular multilevel converter based STATCOM. IEEE Trans Ind Electron 69(3):2166--2179. https://doi.org/10.1109/TIE.2021.3063973
\bibitem{ref_milasi} Milasi RM (2024) A nonlinear adaptive control for a bidirectional DC--AC converter with parameter uncertainties. IEEE Trans Ind Electron 71(8):9551--9558. https://doi.org/10.1109/TIE.2023.3323703
\bibitem{ref_hashem} Hashemzadeh E, et al (2021) A robust control for D-STATCOM under variations of dc-link capacitance. IEEE Trans Power Electron 36(7):8325--8333. https://doi.org/10.1109/TPEL.2020.3026092
\bibitem{ref_ahmed} Ahmed T, et al (2021) Analysis of fractional order sliding mode control in a D-STATCOM integrated power distribution system. IEEE Access 9:70337--70352. https://doi.org/10.1109/ACCESS.2021.3078608
\bibitem{ref_zhou} Zhou X, et al (2021) Control strategy research of D-STATCOM using active disturbance rejection control based on total disturbance error compensation. IEEE Access 9:50138--50150. https://doi.org/10.1109/ACCESS.2021.3069293
\bibitem{ref_gong} Gong C, Sou WK, Lam CS (2023) Reinforcement learning based sliding mode control for a hybrid-STATCOM. IEEE Trans Power Electron 38(6):6795--6800. https://doi.org/10.1109/TPEL.2023.3247835
\bibitem{ref_xiao_mpc} Xiao Q, et al (2022) Modulated model predictive control for multilevel cascaded H-bridge converter-based static synchronous compensator. IEEE Trans Ind Electron 69(2):1091--1102. https://doi.org/10.1109/TIE.2021.3056953
\bibitem{ref_badoni} Badoni M, Singh A, Singh B (2015) Variable forgetting factor recursive least square control algorithm for DSTATCOM. IEEE Trans Power Deliv 30(5):2353--2361. https://doi.org/10.1109/TPWRD.2015.2422139
\bibitem{ref_notch} Singh B, Kant K, Arya SR (2015) Notch filter-based fundamental frequency component extraction to control distribution static compensator for mitigating current-related power quality problems. IET Power Electron 8(9):1758--1766. https://doi.org/10.1049/iet-pel.2014.0486
\bibitem{ref_arya_dg} Arya SR, Niwas R, Bhalla KK, Singh B, Chandra A, Al-Haddad K (2015) Power quality improvement in isolated distributed power generating system using DSTATCOM. IEEE Trans Ind Appl 51(6):4766--4774. https://doi.org/10.1109/TIA.2015.2451093
\bibitem{ref_chan} Channegowda P, John V (2010) Filter optimization for grid interactive voltage source inverters. IEEE Trans Ind Electron 57(12):4106--4114. https://doi.org/10.1109/TIE.2010.2042421
\bibitem{ref_ghoshal_ad} Ghoshal A, John V (2015) Active damping of LCL filter at low switching to resonance frequency ratio. IET Power Electron 8(4):574--582. https://doi.org/10.1049/iet-pel.2014.0355
\bibitem{ref_ghoshal_ctrl} Ghoshal A, John V (2015) A controller design method for 3 phase 4 wire grid connected VSI with LCL filter. S\=adhan\=a 40(5):1481--1499. https://doi.org/10.1007/s12046-015-0393-3
\bibitem{ref_venkat} Venkatramanan D, John V (2020) Dynamic phasor modeling and stability analysis of SRF-PLL-based grid-tie inverter under islanded conditions. IEEE Trans Ind Appl 56(2):1953--1965. https://doi.org/10.1109/TIA.2019.2962763
\bibitem{ref_iec37} IEC (2008) IEC TR 61000-3-7:2008: Electromagnetic compatibility (EMC) --- Part 3-7: Limits --- assessment of emission limits for the connection of fluctuating installations to MV, HV and EHV power systems, 2nd edn. IEC, Geneva
\bibitem{ref_ou} Uhlenbeck GE, Ornstein LS (1930) On the theory of the Brownian motion. Phys Rev 36(5):823--841. https://doi.org/10.1103/PhysRev.36.823
\bibitem{ref_cassie} Cassie AM (1939) Arc rupture and circuit severity: a new theory. CIGRE Report 102, Paris
\bibitem{ref_mayr} Mayr O (1943) Beitr\"age zur Theorie des statischen und des dynamischen Lichtbogens. Arch Elektrotech 37(12):588--608. https://doi.org/10.1007/BF02084317
\bibitem{ref_pso} Kennedy J, Eberhart R (1995) Particle swarm optimization. In: Proc ICNN'95 --- Int Conf Neural Networks, Perth, WA, Australia, vol 4, pp 1942--1948. https://doi.org/10.1109/ICNN.1995.488968
\bibitem{ref_gwo} Mirjalili S, Mirjalili SM, Lewis A (2014) Grey wolf optimizer. Adv Eng Softw 69:46--61. https://doi.org/10.1016/j.advengsoft.2013.12.007
\bibitem{ref_lshade} Tanabe R, Fukunaga AS (2014) Improving the search performance of SHADE using linear population size reduction. In: Proc IEEE Congress on Evolutionary Computation (CEC), Beijing, China, pp 1658--1665. https://doi.org/10.1109/CEC.2014.6900380
\bibitem{ref_embgo} Zhong R, Xu Y, Zhang C, Yu J (2025) Efficient multiplayer battle game optimizer for numerical optimization and adversarial robust neural architecture search. Alex Eng J 113:150--168. https://doi.org/10.1016/j.aej.2024.11.035
\bibitem{ref_rembgo} Chekuri V, Ramarao N, Prashanth NA (2026) REMBGO for evaluation-limited optimization: CEC2022 benchmarking and EAF--STATCOM generalization. arXiv:2610.13471. https://doi.org/10.48550/arXiv.2610.13471
\end{thebibliography}
\end{document}


\setcounter{section}{0}\setcounter{table}{0}\setcounter{figure}{0}\setcounter{equation}{0}
\renewcommand{\thesection}{S\arabic{section}}
\renewcommand{\thetable}{S\arabic{table}}
\renewcommand{\thefigure}{S\arabic{figure}}
\renewcommand{\theequation}{S\arabic{equation}}
\begin{center}{\large\bfseries Supplementary Material}\\[2pt]
\textit{Ensemble-Validated STATCOM Control for EAF Flicker Mitigation}\end{center}
\medskip

\noindent This supplement collects material referenced from the main paper:
plant and arc parameters, the complete optimizer bounds and selected gain
vectors, the per-realization ensemble results, the generator audit,
and secondary time-domain evidence. No result here is new; every table and
figure regenerates from the archived campaign without rerunning any simulation.

\section{Complete symbol list}\label{sms:nomen}
\begin{description}[leftmargin=5.2em,style=nextline,itemsep=0pt,parsep=0pt]
\item[$Q_\mathrm{source}$] Reactive power drawn from the grid at the PCC.
\item[$Q_\mathrm{EAF}$] Reactive power drawn by the furnace.
\item[$Q_\mathrm{STATCOM}$] Reactive power injected by the compensator.
\item[$V_\mathrm{dc}$] Converter dc-link voltage.
\item[$L_1,\,L_2$] Converter-side and grid-side LCL inductances.
\item[$C_f,\,R_d$] LCL filter capacitance and passive damping resistance.
\item[$f_\mathrm{res},\,f_\mathrm{sw},\,f_\mathrm{line}$] LCL resonance, switching and line frequencies.
\item[$Z_b,\,C_b$] Base impedance and base capacitance of the ac side.
\item[$L_g$] Effective interface inductance seen by the current loop.
\item[$e_d,\,e_q$] Direct- and quadrature-axis current-tracking errors.
\item[$i_d,\,i_q$] Direct- and quadrature-axis converter currents.
\item[$i_d^*,\,i_q^*$] Current references from the outer loops.
\item[$K_{p,id},\,K_{i,id}$] DQ-PI $d$-axis current-loop gains.
\item[$K_{p,iq},\,K_{i,iq}$] DQ-PI $q$-axis current-loop gains.
\item[$K_{p,Q},\,T_{iQ}$] DQ-PI reactive-power loop gain and integral time.
\item[$K_{p,vdc},\,K_{i,vdc}$] DQ-PI dc-link loop gains.
\item[$K_{q,p},\,K_{q,i}$] Lyapunov reactive-power loop gains.
\item[$K_{v,p},\,K_{v,i}$] Lyapunov dc-link loop gains.
\item[$K_\mathrm{aw,Q}$] Anti-windup gain of the reactive-power loop.
\item[$K_\mathrm{loss}$] Quadratic converter-loss coefficient.
\item[$I_\mathrm{rated},\,I_\mathrm{trip}$] Rated and trip current levels.
\item[$I_{q,\max}$] Reactive-current limit of the radial limiter.
\item[$X,\,R$] Grid reactance and resistance at the PCC.
\item[$S_\mathrm{rated},\,V_{LL}$] Converter rating and line-to-line voltage.
\end{description}

\section{LCL filter sizing and attenuation}\label{sms:lcl}
The converter is connected to the grid through an LCL filter designed according to 
the standard procedure in. The design sequence comprises defining the base 
quantities, selecting the filter capacitance, determining the converter- and grid-side 
inductances, setting the resonance frequency, and incorporating passive damping. With the
rated ac-side base impedance $Z_b=V_{LL}^{2}/S_\mathrm{rated}=415^{2}/500\,000=0.344~\Omega$
and base capacitance $C_b=1/(\omega_0 Z_b)=9241~\mu$F
($\omega_0=2\pi f_\mathrm{line}$), the filter capacitor is
kept under 10\% of $C_b$, giving $C_f=280~\mu$F ($3.0\%$ of $C_b$). The
dc-derived quantity $V_\mathrm{dc}/2=600$~V enters only the modulation limit
$V_\mathrm{pk,max}$, not the base impedance. The inverter-side
inductance $L_1=0.22$~mH is set by the ripple-current limit
$L_1=V_\mathrm{dc}/(8\,\Delta I_\mathrm{max}f_\mathrm{sw})$, where
$\Delta I_\mathrm{max}$ is the allowed peak-to-peak ripple; the chosen value
gives a ripple amplitude of $\pm68$~A, i.e., 9.8\% of $I_\mathrm{rated}$, and the grid-side
inductance is $L_2=0.20$~mH ($L_2/L_1=0.91$). The resonance frequency, from
equation~\eqref{eq:lclres},
\begin{equation}
  f_\mathrm{res}=\frac{1}{2\pi}\sqrt{\frac{L_1+L_2}{L_1 L_2 C_f}}\approx 930~\mathrm{Hz},
  \label{eq:lclres}
\end{equation}
satisfies $10f_\mathrm{line}<f_\mathrm{res}<f_\mathrm{sw}/2$. The undamped
voltage-to-current transfer function of the filter branch is third order, as in
equation~\eqref{eq:lcltf},
\begin{equation}
  G_\mathrm{LCL}(s)=\frac{i_{2}(s)}{v_{i}(s)}=\frac{1}{s^{3}L_1 L_2 C_f+s\,(L_1+L_2)},
  \label{eq:lcltf}
\end{equation}
so for $\omega\gg\omega_\mathrm{res}$ the magnitude rolls off at $-60$~dB/decade,
three times steeper than a single-$L$ filter, which suppresses the
switching-frequency current ripple. The selected filter attenuates
the 5~kHz switching component by 34.8~dB with
$f_\mathrm{sw}/f_\mathrm{res}=5.38$, exceeding the adopted 30~dB attenuation
criterion; a
passive damping resistor $R_d\approx1/(3\omega_\mathrm{res}C_f)=0.2~\Omega$ damps
the resonant peak. Active damping would remove this loss
element, but with $f_\mathrm{sw}/f_\mathrm{res}=5.38$ the ratio is high enough
that the simple passive resistor is adequate here.

\section{Objective-function definition and tuning protocol}\label{sms:obj}
\noindent The main paper states the objective and the common-budget protocol. The complete feature definitions, flicker-band filters, moving-window construction, penalty formulae and selection pipeline are reproduced here unchanged from the original campaign specification.

The Lyapunov and the DQ-PI controllers are tuned with a common performance
score evaluated over the post-transient window $t\in[0.50,1.00]$~s. The score
combines a normalized reactive-power term $T_Q$ and a normalized voltage term
$T_V$,
\begin{equation}
  T_Q=\frac{Q_\mathrm{fluc}}{Q_\mathrm{fluc}^\mathrm{ref}},\qquad
  T_V=\frac{V_\mathrm{pcc}}{V_\mathrm{pcc}^\mathrm{ref}},
  \label{eq:TQTV}
\end{equation}
augmented by graded quality and safety penalties,
\begin{equation}
  J_\mathrm{perf}=0.60\,T_Q+0.40\,T_V,\qquad
  J_\mathrm{tot}=J_\mathrm{perf}+P_{Q}+P_{V_\mathrm{dc}}+P_\mathrm{safety}.
  \label{eq:J}
\end{equation}
This score drives the gain search only; it is not a flicker index. The two raw
features in Eq.~\eqref{eq:TQTV} are defined exactly as follows, both evaluated
over the performance window. With $Q$ the source reactive power (kVAr) and
$Q_d=\mathrm{detrend}(Q)$ its linearly detrended value,
\begin{equation}
  Q_\mathrm{fluc}=1.5\,\mathrm{std}(Q_d)
  +2.0\,\mathrm{rms}\!\left[\mathrm{BPF}_{5\text{--}15}(Q_d)\right],
  \label{eq:qfluc}
\end{equation}
where $\mathrm{BPF}_{5\text{--}15}$ is a zero-phase (forward--backward)
second-order Butterworth band-pass over $5$--$15$~Hz. For the voltage feature,
each PCC phase voltage is reduced to its one-cycle ($50$~Hz) sliding-RMS envelope
$e$, expressed as the relative deviation $\tilde v=(e-\bar e)/\bar e$ from its
in-window mean, then decimated to $500$~Hz and linearly detrended to give
$v$; the per-phase composite is
\begin{equation}
  c=2.0\,\mathrm{rms}(v_\mathrm{lp})
  +3.0\,\mathrm{rms}\!\left[\mathrm{BPF}_{7\text{--}12}(v)\right]
  +\mathrm{mean}\!\left[\mathrm{movstd}(v_\mathrm{lp},0.1~\mathrm{s})\right],
  \label{eq:vpcc}
\end{equation}
where $v_\mathrm{lp}$ is $v$ low-pass filtered at $35$~Hz and both voltage filters
are zero-phase second-order Butterworth, and $V_\mathrm{pcc}$ is the maximum of
$c$ over the three phases. Both terms share one common
deterministic uncompensated reference ($Q_\mathrm{fluc}^\mathrm{ref}$,
$V_\mathrm{pcc}^\mathrm{ref}$), so the normalization denominators are identical
for the two arc modes; a mode-specific reference would partially divide out the
greater absolute severity of the stochastic arc. $J_\mathrm{perf}$ is therefore
a common-scale cross-mode severity score, not a percentage mitigation relative
to each mode's own uncompensated baseline. The graded penalties act only near
the operating limits:
$P_{Q}=0.05\,[\max(0,(|\overline{Q}|-15)/15)]^{2}$ penalizes a window-mean
source reactive power $\overline{Q}$ above $15$~kVAr,
$P_{V_\mathrm{dc}}=0.03\,[\max(0,(\Delta V_\mathrm{dc,pk}-0.03)/0.02)]^{2}$
penalizes a dc-link peak deviation above $3\%$, and $P_\mathrm{safety}$ collects
soft converter-saturation and current-tracking costs together with hard
feasibility gates on reactive power and dc-link excursion. The same score and weights apply to
both controllers, so $J_\mathrm{tot}$ is directly comparable between them, and its
internal feasibility flag is a converter-limit check rather than an IEC
compliance statement. The 600~s IEC \pst{} is obtained separately
(Section~VII of the main paper). The term-by-term normalization is given above;
an optimizer-level statistical comparison is outside the scope of this paper.

Five metaheuristics generate the tuned six-gain vectors: particle swarm
optimization (PSO), the grey wolf optimizer (GWO),
a fixed-population SHADE variant (FP-SHADE), EMBGO, and its refined variant
REMBGO. FP-SHADE uses the success-history adaptation and
current-to-$p$best/1 archive of L-SHADE but keeps a fixed
population of $N=40$ without linear population-size reduction, so the cited
reference is the source of the adaptation mechanism rather than evidence for
canonical L-SHADE. Each controller is tuned in both arc modes, giving
twenty optimization tasks (five optimizers $\times$ two controllers $\times$ two
arc modes); each task uses ten optimizer seeds, for 200 optimization runs in
total. Every run uses a population of $N=40$ over 25 update iterations, giving
$1040$ objective evaluations including the initial population
($40+40\times25=1040$), and follows the matched tuning protocol
specified for the campaign: the validated warm-start
vector $\symbf{x}_0$ is used as the first population member, three bounded
multiplicative neighbours are added (independent coordinate factors drawn from
$\mathcal{U}[0.8,1.2]$), and the remaining members are filled by Latin-hypercube
sampling within the prescribed search box under a controlled run-specific seed.
Within a given optimizer--controller--arc task the five optimizers share the same
integer seed for a given run index, so the ten seeds index optimizer randomness
under a matched initialization rather than independent experimental conditions.
The DQ-PI search uses mode-specific bounds (Table~\ref{tab:bounds}); the
stochastic-mode bounds were prospectively shifted to the validated stable search
region, including a higher range for the slow reactive-integral action. The
Lyapunov search uses a single validated global bound set. The final champion gain
sets are listed in Table~\ref{tab:gains}. The gain-set selection and evaluation
follow the three-stage pipeline of Fig.~2 of the main paper. Each optimization run
is conducted against $J_\mathrm{tot}$ and produces one run-level optimized gain
set. In the first selection stage, within each optimizer--controller--arc task the
seed with the lowest \pscr{} is chosen for 600-s evaluation, giving twenty
task-wise selected vectors (five optimizers
$\times$ two controllers $\times$ two arc modes). In the second stage these twenty
vectors are simulated for 600~s, and for each controller--arc-mode combination the
headline champion is the vector with the lowest 600~s \pst{} among its five
optimizer-task selections. The two indices thus perform two distinct selection
roles, and the same metric is never used for both. In this control paper the
optimizers are only gain-generation tools; a statistical comparison between them
is outside its scope.

\FloatBarrier

\section{Corrected-generator audit}\label{sms:rng}
A post-campaign audit of the pseudo-random generator used for the arc-length
perturbation established that the recursion had been evaluated in
double precision, so that the product exceeded the exactly representable integer
range and the low-order bits were lost. The generator consequently entered a
short cycle, and the stochastic realization used throughout the original campaign
repeated with a period of $10\,466$ solver steps ($104.66$~ms, $9.55$~Hz) rather
than providing a broadband perturbation. The executed perturbation therefore had
a mean of $-0.055$ and a standard deviation of $0.003$, against the intended zero
mean and $\sigma_n=0.015$. An exact 64-bit integer implementation of the same
recursion removes the truncation; over the arc seeds audited here it reproduces
the intended statistics (grand mean $-9.0\times10^{-5}$, mean standard deviation
$0.0147$). The deterministic arc mode is bit-identical between the two
implementations and is therefore unaffected. Because this was discovered after
the optimization campaigns had been completed and frozen, the gain sets were
\emph{not} retuned; instead, Section~VI-D of the main paper reports a fixed-gain
validation of the already-selected controllers on independent stochastic
realizations. All results reported as single-trajectory stochastic values in
Supplementary Table~S5--Table~3 of the main paper refer to the STO-SR stochastic-mode realization (STO-SR) and are retained for provenance and comparison.

\section{Extended stability derivation}\label{sms:proof}
\noindent Differentiating the extended storage function of the main paper along
the closed-loop error dynamics gives, term by term,
\[
\dot V=\sum_x e_x\Bigl(-k_{1,i}e_x-n_x-\tfrac{k_{i,i}}{L_g}\xi_{ix}+\tilde d_x\Bigr)
       +\tfrac{k_{i,i}}{L_g}\sum_x \xi_{ix}\dot\xi_{ix},
\]
and substituting $\dot\xi_{ix}=e_x$ makes the two $\left(k_{i,i}/L_g\right)e_x\xi_{ix}$
terms cancel identically, leaving
$\dot V=-k_{1,i}\lVert\symbf e\rVert^{2}-\sum_x e_xn_x+\symbf e^{\top}\tilde{\symbf d}$.
With $\sigma_x=e_x/(|e_x|+\phi_i)$ the nonlinear term satisfies
$-e_xn_x=-k_{2,i}e_x^{2}(|e_x|+\epsilon)^{\alpha}/(|e_x|+\phi_i)\le0$ for every
real $e_x$ and $k_{2,i}>0$, with equality only at $e_x=0$. Setting
$\tilde{\symbf d}=\symbf 0$ therefore gives a negative semidefinite $\dot V$ whose
zero set is $\symbf e=\symbf 0$; on that set $n_x=0$ and $\dot e_x=0$, so the
closed-loop dynamics force $\xi_{ix}=0$ whenever $k_{i,i}>0$.

\section{Plant, arc and controller parameters}\label{sm:params}
\begin{table}[!htbp]
\centering
\caption{Stochastic EAF-model parameters (Cassie--Mayr arc with reproducible correlated arc-length perturbation).}
\label{tab:eafparams}
\setlength\tabcolsep{6pt}\renewcommand{\arraystretch}{1.05}\small
\begin{tabular}{lll}
\toprule
\textbf{Parameter} & \textbf{Symbol} & \textbf{Value}\\
\midrule
Flicker modulation depth   & $m_\mathrm{flicker}$ & 0.065\\
Correlated-noise time const. & $\tau_n$ & 0.25~s\\
Noise intensity            & $\sigma_n$ & 0.015\\
Arc-length base (clipped)  & $L_0$ & 1.0, to $[0.6,1.4]$\\
Per-phase length offsets   & -- & $0,\,0.01,\,0.02$\\
Phase noise seeds          & -- & 59, 76, 93\\
Cassie voltage constant    & $V_{c0}$ & 200~V\\
Mayr power constant        & $P_{m0}$ & $5\times10^{4}$~W\\
Cassie / Mayr time const.  & $\tau_c,\tau_m$ & 0.5~ms, 0.3~ms\\
Re-ignition voltage        & $V_\mathrm{re}$ & 280~V\\
Extinction current         & $i_\mathrm{ext}$ & 2~A\\
Extinction rate coefficient & $p_\mathrm{ext}$ & 0.003~ms$^{-1}$\\
Re-ignition conductance    & $g_\mathrm{re}$ & 0.25~S\\
Arc series resistance      & $r_a$ & 0.02~$\Omega$\\
Solver step                & $T_s$ & 10~$\mu$s\\
\bottomrule
\end{tabular}
\end{table}
\begin{table*}[!htbp]
\centering
\caption{Final controller performance summary, comparing the uncompensated
system with the selected 600~s champion of each controller. The \pst{} values
use the 600~s IEC short-term interval; the remaining metrics are computed over
the same 600~s evaluation run. Bold identifies the best compensated value within
each EAF mode and metric. $Q_\sigma$ is the fluctuating (detrended) source
reactive power.}
\label{tab:p1-controller-summary}
\setlength\tabcolsep{5pt}\renewcommand{\arraystretch}{1.12}\small
\begin{tabular}{llcccccc}
\toprule
\textbf{EAF} & \textbf{Operating condition} &
$\symbf{P}_{\mathrm{st}}\downarrow$ &
$\symbf{Q}_{\sigma}$ (kVAr)$\downarrow$ &
\textbf{P--Q PF}$\uparrow$ &
$\symbf{V}_{\mathrm{dc}}$ \textbf{dev. (\%)}$\downarrow$ &
$\overline{\symbf{\mathrm{THD}}}_{V}$ \textbf{(\%)}$\downarrow$ &
\textbf{Norm. IAE}$\downarrow$\\
\midrule
Det. & Uncompensated      & 1.812 & 8.73 & 0.896 & -- & 0.796 & -- \\
Det. & DQ-PI optimized    & 0.680 & 2.23 & 0.999 & 4.98 & 1.80 & \textbf{0.0339} \\
Det. & Lyapunov optimized & \textbf{0.635} & \textbf{1.18} & \textbf{1.000} & \textbf{0.205} & \textbf{1.63} & 0.0438 \\
\midrule
STO-SR & Uncompensated      & 2.262 & 10.88 & 0.799 & -- & 0.874 & -- \\
STO-SR & DQ-PI optimized    & \textbf{0.800} & 2.04 & 0.974 & 5.98 & 1.93 & \textbf{0.0352} \\
STO-SR & Lyapunov optimized & 0.889 & \textbf{1.70} & \textbf{1.000} & \textbf{0.266} & \textbf{0.889} & 0.0569 \\
\bottomrule
\end{tabular}
\vspace{2pt}
{\footnotesize\begin{flushleft}
The P--Q PF is $|\overline P|/\sqrt{\overline P^2+\overline Q^2}$ and is not a
true waveform power factor. Voltage THD is the mean of phases a, b, and c
through the 50th harmonic; the uncompensated value is lowest because converter
switching adds a modest harmonic component. The IAE is the normalized sum of the
$d$- and $q$-axis current-tracking integral absolute errors. dc-link deviation
and IAE are not applicable when the STATCOM is off. Saturation is 0\% for all
compensated cases.
\end{flushleft}}
\end{table*}

\section{Optimizer bounds and selected gain vectors}\label{sm:gains}
\begin{table}[!htbp]
\centering
\caption{Optimizer search bounds. The DQ-PI bounds are mode-specific; the
Lyapunov bounds are a single validated global set used for both arc modes.}
\label{tab:bounds}
\setlength\tabcolsep{4pt}\renewcommand{\arraystretch}{1.1}\small
\begin{tabular}{lcccccc}
\toprule
\multicolumn{7}{l}{\textit{DQ-PI}}\\
\textbf{Bound} & $K_{p,id}$ & $K_{i,id}$ & $K_{p,iq}$ & $K_{i,iq}$ & $K_{p,Q}$ & $T_{iQ}$ (s)\\
\midrule
Det. lower & 0.15 & 30  & 1.20 & 260 & $-38$ & 15\\
Det. upper & 0.30 & 100 & 1.50 & 370 & $-28$ & 23\\
STO-SR lower & 0.08 & 20  & 1.25 & 150 & $-45$ & 5\\
STO-SR upper & 0.35 & 220 & 1.45 & 650 & $-25$ & 20\\
\midrule
\multicolumn{7}{l}{\textit{Lyapunov} (global, both modes)}\\
\textbf{Bound} & $K_\mathrm{droop}$ & $K_{q,p}$ & $K_{q,i}$ & $k_{1,i}$ & $k_{i,i}$ & $k_{2,i}$\\
\midrule
Lower & 0     & 0.5 & 100  & 50   & 5   & 300\\
Upper & 60000 & 20  & 3000 & 4500 & 600 & 3000\\
\bottomrule
\end{tabular}
\end{table}
\begin{table}[!htbp]
\centering
\caption{Final champion gain sets (best 600~s case per controller and arc mode).
Each gain is listed exactly as implemented in the controller laws of
Section~V of the main paper; explicit unit labels are omitted because the
nonlinear inner-law scaling and internal reactive-power normalization make
compact unit annotation ambiguous, and each gain's dimension follows directly
from its defining equation.}
\label{tab:gains}
\setlength\tabcolsep{4pt}\renewcommand{\arraystretch}{1.1}\small
\begin{tabular}{lcccccc}
\toprule
\multicolumn{7}{l}{\textit{DQ-PI}}\\
\textbf{EAF (opt.)} & $K_{p,id}$ & $K_{i,id}$ & $K_{p,iq}$ & $K_{i,iq}$ & $K_{p,Q}$ & $T_{iQ}$ (s)\\
\midrule
Det. (GWO)      & 0.300 & 100.0 & 1.317 & 370.0 & $-28.00$ & 15.00\\
STO-SR (FP-SHADE) & 0.285 & 133.4 & 1.297 & 630.5 & $-28.72$ & 14.15\\
\midrule
\multicolumn{7}{l}{\textit{Lyapunov}}\\
\textbf{EAF (opt.)} & $K_\mathrm{droop}$ & $K_{q,p}$ & $K_{q,i}$ & $k_{1,i}$ & $k_{i,i}$ & $k_{2,i}$\\
\midrule
Det. (EMBGO) & 0    & 8.059 & 2948 & 2279 & 5.00  & 2807\\
STO-SR (EMBGO) & 7421 & 6.921 & 3000 & 3591 & 6.731 & 1199\\
\bottomrule
\end{tabular}
\end{table}

\section{Evaluation windows and per-realization results}\label{sm:results}
\begin{table}[!htbp]
\centering
\caption{Cumulative flicker-index convergence with observation duration,
converging to the 600~s IEC short-term value for the deterministic and
stochastic EAF models. The optimized cases are the selected champion gain sets
of Table~3 of the main paper (DET DQ-PI: GWO; DET Lyapunov:
EMBGO; STO-SR DQ-PI: FP-SHADE; STO-SR Lyapunov: EMBGO). The 20~s column
is the cumulative estimate from the corresponding 600~s evaluation replay and
should not be confused with the separate campaign screening quantity \pscr{}.}
\label{tab:windows}

\renewcommand{\arraystretch}{1.12}
\setlength{\tabcolsep}{7.2pt}
\small

\begin{tabular}{lcccccc}
\toprule
\textbf{Case}
& \textbf{20 s}
& \textbf{50 s}
& \textbf{100 s}
& \textbf{200 s}
& \textbf{300 s}
& \textbf{600 s} \\
\midrule
Uncompensated (Det.)      & 1.7786 & 1.7950 & 1.8053 & 1.8100 & 1.8112 & 1.8122 \\
DQ-PI optimized (Det.)    & 0.6560 & 0.6833 & 0.6744 & 0.6774 & 0.6794 & 0.6796 \\
Lyapunov optimized (Det.) & 0.6350 & 0.6484 & 0.6479 & 0.6408 & 0.6376 & 0.6346 \\
\midrule
Uncompensated (STO-SR)      & 2.2102 & 2.2348 & 2.2484 & 2.2578 & 2.2601 & 2.2623 \\
DQ-PI optimized (STO-SR)    & 0.8006 & 0.8083 & 0.7998 & 0.7944 & 0.8010 & 0.7999 \\
Lyapunov optimized (STO-SR) & 0.8651 & 0.8844 & 0.8759 & 0.8841 & 0.8905 & 0.8892 \\
\bottomrule
\end{tabular}
\end{table}
\begin{table}[!htbp]
\centering
\caption{Controller-level 20~s screening statistics of the flicker screening
index (\pscr{}), pooled over the five optimizers and ten seeds of each
controller--EAF combination ($n=50$ observations per row). The seeds are
optimizer initialization seeds under one reproducible arc trajectory, not
arc-noise realizations. The 20~s values are used only for comparative gain-set
screening and are not numerically interchangeable with the 600~s \pst{} values
of Table~3 of the main paper.}
\label{tab:tenseed}
\setlength\tabcolsep{8pt}\renewcommand{\arraystretch}{1.08}\small
\begin{tabular}{lcccc}
\toprule
\textbf{Scenario} & \textbf{Mean$\pm$std} & \textbf{Min} & \textbf{Max} & $n$\\
\midrule
DET DQ-PI    & $0.743\pm0.075$ & 0.638 & 0.952 & 50\\
DET Lyapunov & $0.668\pm0.036$ & 0.625 & 0.762 & 50\\
STO-SR DQ-PI       & $0.899\pm0.090$ & 0.768 & 1.179 & 50\\
STO-SR Lyapunov    & $0.910\pm0.030$ & 0.865 & 0.999 & 50\\
\bottomrule
\end{tabular}
\end{table}

\section{Five-champion robustness summary}\label{sm:robust}
\begin{table}[!htbp]
\centering
\caption{Controller-level 600~s \pst{} across the five optimizer-task champions
of each controller--arc-mode combination ($n=5$; descriptive summary, not an
optimizer ranking). Bold marks the lower controller mean within each arc mode.}
\label{tab:robust}
\setlength\tabcolsep{6pt}\renewcommand{\arraystretch}{1.1}\small
\begin{tabular}{llccc}
\toprule
\textbf{Controller} & \textbf{EAF} & \textbf{Mean} & \textbf{Std} & \textbf{Best}\\
\midrule
DQ-PI    & Det. & 0.701 & 0.023 & 0.680\\
Lyapunov & Det. & \textbf{0.647} & 0.009 & \textbf{0.635}\\
DQ-PI    & STO-SR & \textbf{0.813} & 0.011 & \textbf{0.800}\\
Lyapunov & STO-SR & 0.906 & 0.016 & 0.889\\
\bottomrule
\end{tabular}
\end{table}

\section{Secondary time-domain evidence}\label{sm:timedomain}

\begin{figure}[H]
  \centering
  \begin{subfigure}[t]{0.48\textwidth}
    \includegraphics[width=\linewidth]{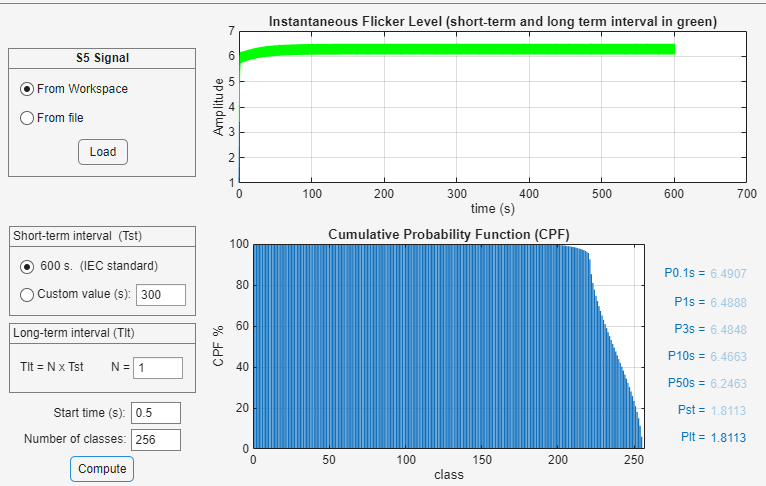}
    \caption{Uncompensated deterministic case: analyzer $\pst{}=1.8113$
    (exported value $1.8122$).}
    \label{fig:fm_uncomp}
  \end{subfigure}\hfill
  \begin{subfigure}[t]{0.48\textwidth}
    \includegraphics[width=\linewidth]{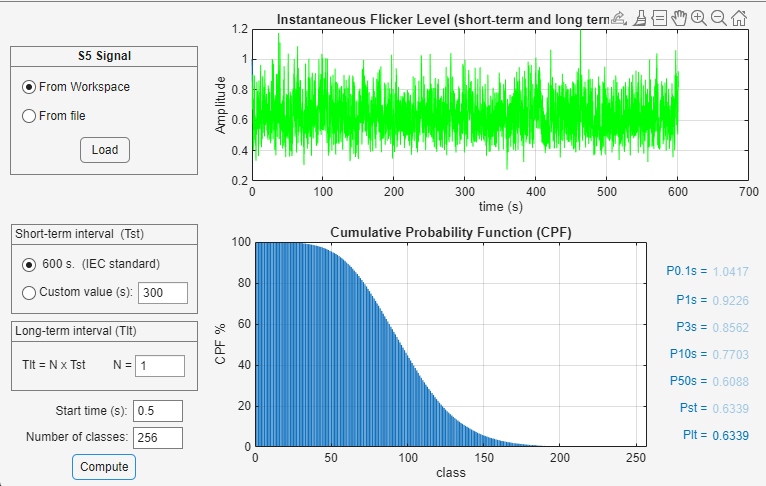}
    \caption{Deterministic EMBGO-tuned Lyapunov case: analyzer
    $\pst{}=0.6339$ (exported value $0.6346$).}
    \label{fig:fm_lyap}
  \end{subfigure}
  \caption{Digital flickermeter analyzer outputs for the deterministic EAF over
  the 600~s IEC short-term interval (start time 0.5~s, 256 classes):
  (a) uncompensated operation and (b) the optimized EMBGO-tuned Lyapunov gain
  set. Each panel shows the instantaneous flicker level, the cumulative
  probability function, and the percentile quantities used to obtain \pst{}. The
  analyzer display values differ from the exported values by less than $0.001$;
  the exported values are used for all numerical comparisons in the paper.}
  \label{fig:flickermeter}
\end{figure}

\begin{figure}[!htbp]
  \centering
  \includegraphics[width=\linewidth]{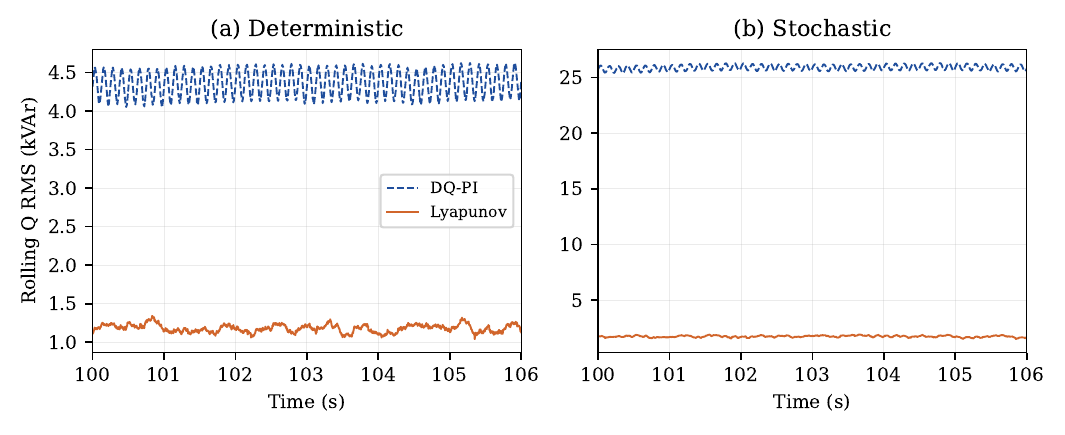}
  \caption{Rolling source-reactive-power RMS (0.2~s trailing window) for the
  optimized DQ-PI and Lyapunov champions under deterministic and stochastic
  operation, over the 600~s evaluation run.}
  \label{fig:qrms_nominal_opt}
\end{figure}

\begin{figure}[!htbp]
  \centering
  \includegraphics[width=\linewidth]{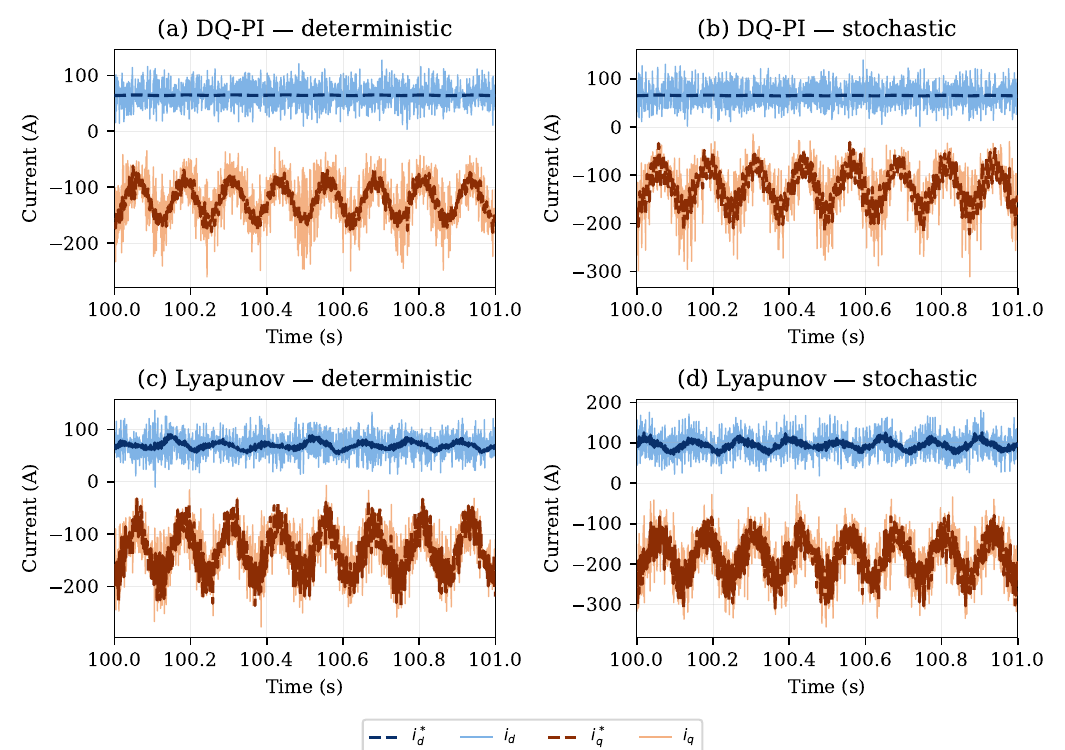}
  \caption{Reference and measured $d$- and $q$-axis currents (blue: $d$-axis,
  orange: $q$-axis; dashed: reference, solid: measured) under deterministic and
  stochastic EAF operation, over a representative 1~s steady-state segment.}
  \label{fig:tracking_combined}
\end{figure}
\begin{figure}[!htbp]
  \centering
  \includegraphics[width=\linewidth]{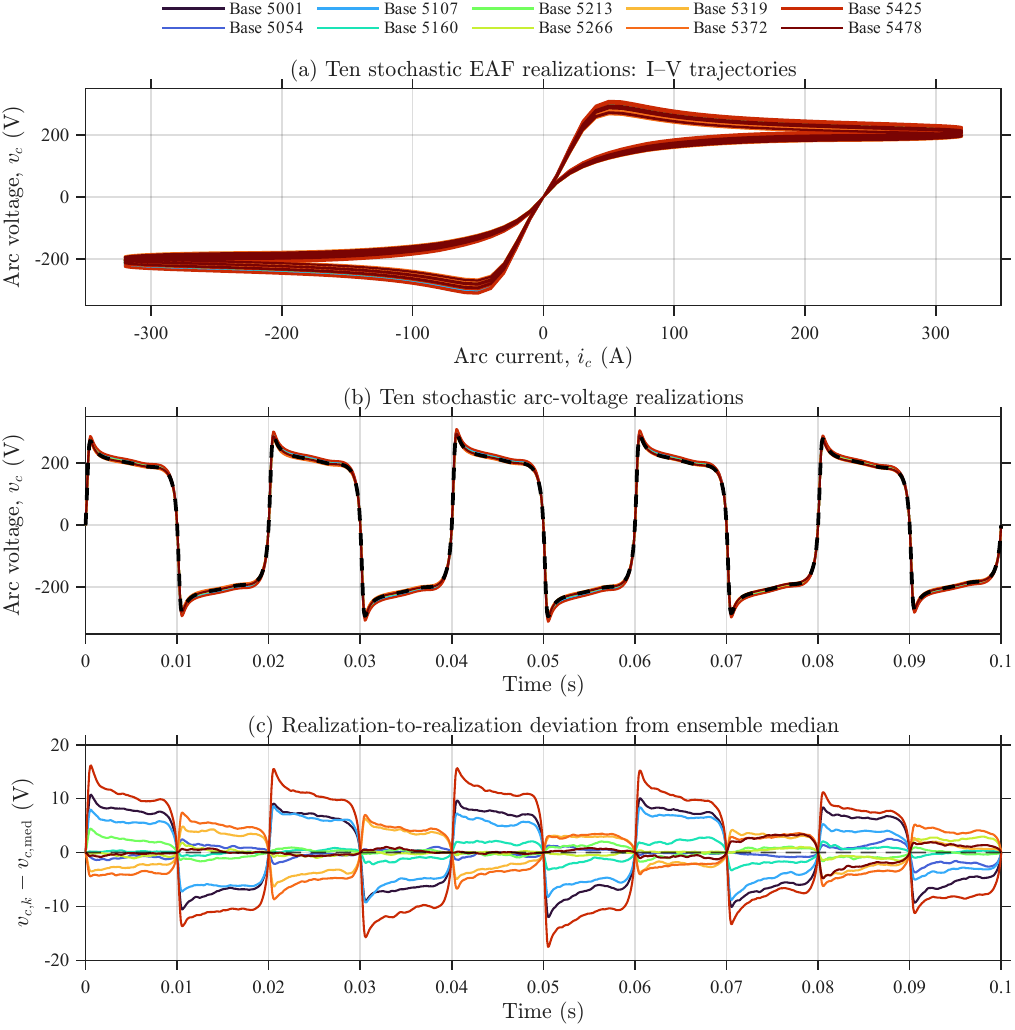}
  \caption{Per-realization structure of the stochastic ensemble, over five line
  cycles. (a) Arc I--V trajectories and (b) arc voltages for the ten
  realizations, one colour per base seed; (c) each realization minus the
  pointwise ensemble median. The ten coloured traces are ten \emph{separate}
  model realizations evaluated independently, not ten simultaneous arcs inside
  one simulation; within a realization the three phase seeds follow the frozen
  rule $s_\phi = s_\mathrm{base} + 17\,\mathrm{phase\_id}$. Panels (a) and (b)
  are dominated by the common nonlinear arc waveform, which is identical by
  construction, so the realizations overlie one another. Panel (c) removes that
  common component and is therefore the diagnostic panel: it resolves the
  seed-dependent dispersion, of order $\pm15$~V, that the ensemble contributes.
  The pointwise median subtracted in (c) is post-processing and is not itself a
  physical realization. These are $0.1$~s intrinsic arc-voltage records and
  cannot be used to infer \pst{}; flicker severity is obtained only from the
  600~s IEC evaluations of Section~\ref{sm:results}.}
  \label{fig:eaf10}
\end{figure}